\documentclass[12pt, draftclsnofoot, onecolumn]{IEEEtran}
\usepackage{hyperref}
\usepackage{setspace}
\usepackage{lettrine}
\usepackage{epsfig}
\usepackage{xurl}
\usepackage{longtable}
\usepackage[linesnumbered,boxed,ruled,commentsnumbered]{algorithm2e}
\usepackage{footnote}
\usepackage{cite}
\usepackage{float}
\usepackage{booktabs}
\usepackage{multicol}
\usepackage{dblfloatfix}
\usepackage{amssymb}
\usepackage{amsmath}
\usepackage{amsthm}
\usepackage{comment}
\usepackage{lscape}
\usepackage{multirow}
\usepackage{amsmath,mathabx}
\usepackage{color}
\usepackage{amsfonts}
\usepackage{tabularx,tabulary}
\usepackage[font=footnotesize]{caption}
\usepackage{graphicx}
\usepackage{dblfloatfix}    % To enable figures at the bottom of page
\usepackage{bbm}
\usepackage{bigints}
\usepackage[usestackEOL]{stackengine}
\usepackage{color}
\usepackage{multicol}
\usepackage{multirow}
\usepackage{caption}
\usepackage{subcaption}
\usepackage{tabularx}
\usepackage[T1]{fontenc}
\usepackage[shortlabels]{enumitem}

\usepackage{verbatim} % for annotation

\usepackage{scalerel}
\usepackage{tikz}
\usetikzlibrary{svg.path}
\definecolor{orcidlogocol}{HTML}{A6CE39}
\tikzset{
  orcidlogo/.pic={
    \fill[orcidlogocol] svg{M256,128c0,70.7-57.3,128-128,128C57.3,256,0,198.7,0,128C0,57.3,57.3,0,128,0C198.7,0,256,57.3,256,128z};
    \fill[white] svg{M86.3,186.2H70.9V79.1h15.4v48.4V186.2z}
                 svg{M108.9,79.1h41.6c39.6,0,57,28.3,57,53.6c0,27.5-21.5,53.6-56.8,53.6h-41.8V79.1z M124.3,172.4h24.5c34.9,0,42.9-26.5,42.9-39.7c0-21.5-13.7-39.7-43.7-39.7h-23.7V172.4z}
                 svg{M88.7,56.8c0,5.5-4.5,10.1-10.1,10.1c-5.6,0-10.1-4.6-10.1-10.1c0-5.6,4.5-10.1,10.1-10.1C84.2,46.7,88.7,51.3,88.7,56.8z};
  }
}

\newcommand\orcidicon[1]{\href{https://orcid.org/#1}{\mbox{\scalerel*{
\begin{tikzpicture}[yscale=-1,transform shape]
\pic{orcidlogo};
\end{tikzpicture}
}{|}}}}

\newcommand{\norm}[1]{\left\lVert#1\right\rVert}

\begin{document}

\title{Data Center-Enabled High Altitude Platforms: A Green Computing Alternative   }
\author{Wiem Abderrahim \orcidicon{0000-0001-5896-2307}, Osama Amin \orcidicon{0000-0002-0026-5960} and Basem Shihada \orcidicon{0000-0003-4434-4334}\\
\thanks{The authors are with the  Computer, Electrical and Mathematical Sciences and Engineering (CEMSE) Divison, King Abdullah University of Science and Technology (KAUST), Thuwal 23955, Makkah Prov., Saudi Arabia (e-mail: { wiem.abderrahim, osama.amin, basem.shihada }@kaust.edu.sa)}}
\maketitle

\begin{abstract}
Information technology organizations and companies are seeking greener alternatives to traditional terrestrial data centers to mitigate global warming and reduce carbon emissions. Currently, terrestrial data centers consume a significant amount of energy, estimated at about $1.5\%$ of worldwide electricity use. Furthermore, the increasing demand for data-intensive applications is expected to raise energy consumption, making it crucial to consider sustainable computing paradigms. In this study, we propose a data center-enabled High Altitude Platform (HAP) system, where a flying data center supports the operation of terrestrial data centers. We conduct a detailed analytical study to assess the energy benefits and communication requirements of this approach. Our findings demonstrate that a data center-enabled HAP is more energy-efficient than a traditional terrestrial data center, owing to the naturally low temperature in the stratosphere and the ability to harvest solar energy. Adopting a data center-HAP can save up to $14\%$ of energy requirements while overcoming the offloading outage problem and the associated delay resulting from server distribution. Our study highlights the potential of a data center-enabled HAP system as a sustainable computing solution to meet the growing energy demands and reduce carbon footprint.

\end{abstract}


\newpage
\section{Introduction}
Data centers are one of the top enabling technologies for the information technology industry, with a global market spending of  216,095 million US dollars in 2022 \cite{hogade2021energy,gart2023}. They are considered the mission-critical infrastructure of this computing era because they play a pivotal role in processing and storing our continuously-growing data \cite{dayarathna2015}. This role is unprecedentedly crucial given the data-intensive applications used in nowadays’ advanced fields such as artificial intelligence and internet-of-everything  applications  \cite{hogade2021energy,kaur2019big, aslani2020rethinking, abderrahim2022proactive}, which require extensive growth of this infrastructure’s sizes and functionalities \cite{hogade2021energy,dukic2020beyond}. However, data centers face major energy-efficiency issues \cite{dayarathna2015,wan2020sustainability} as a large-scale computing infrastructure. Indeed, they consume not only substantial amounts of energy, around 1.5\% of the worldwide electricity use but also the annual growth of the consumed energy by data centers is predicted to rise exponentially in the upcoming years \cite{berezovskaya2020modular,li2017holistic,hogade2021energy,hu2019joint,albers2019energy}. For example, the electricity demand of Google’s data centers increased twenty fold over the last ten years \cite{li2017holistic}. Moreover, China’s data centers will be devouring more than 400 billion kWh by 2030, accounting for 3.7 percent of the country's total electricity consumption \cite{china}. In addition, the annual energy cost of data centers is expected to surpass their construction cost and equipment price within the upcoming few years \cite{dayarathna2015,li2017holistic,hogade2021energy}. These statistics are particularly alarming because the non-renewable energy is still the predominant source to generate electricity nowadays \cite{berezovskaya2020modular}. Therefore, serious research efforts should be conducted towards finding practical solutions that improve the current data center energy efficiency, which can help reduce global carbon emissions. According to Gartner, Inc., it is expected that by 2027, around 75\% of organizations will have implemented a data center infrastructure sustainability program due to cost optimization and stakeholder pressures. %This represents a significant increase from $5\%$ of organizations who have such a program in place in 2022
\cite{gart2023_May}.  

\subsection{Literature Review}
Data center energy consumption is primarily distributed between the cooling infrastructure (30\%-40\%) and the computing infrastructure (26\%-56\%) \cite{dayarathna2015,berezovskaya2020modular,ismail2020computing}. As a result, various energy-aware approaches have been investigated in the literature to improve data center energy efficiency from both computation and cooling perspectives \cite{chan2018optimal,berezovskaya2020modular,li2017holistic,zhou2015carbon,hogade2018minimizing,hogade2021energy,wierman2009power,liu2014greening}.
On the one hand, cooling energy can be reduced through strategies such as raised floors, racks' arrangement following hot/cold aisles, chillers configuration, and fan optimization \cite{chan2018optimal,berezovskaya2020modular}. Additionally, load balancing across geographically distributed data centers can decrease cooling power. For instance, workload distribution and scheduling should consider the energy/cooling efficiency of servers, as well as electricity costs and experienced delays \cite{li2017holistic,zhou2015carbon,hogade2018minimizing,hogade2021energy}. 
On the other hand, data center computational energy can be reduced through dynamic management of capacity by controlling idle servers \cite{dayarathna2015,zhou2015carbon,yao2018ts,albers2019energy}. Furthermore, optimizing server speed by adjusting central processing unit (CPU) frequency can help reduce associated computational power \cite{wierman2009power}.
Another way to decrease energy consumption in data centers is to leverage green options. For example, renewable energy sources like wind and solar power can supply electricity to data centers, as demonstrated by companies such as Apple, Google, Microsoft, and Facebook \cite{hogade2018minimizing,agarwal2021redesigning}. Consequently, it is vital to explore how data centers can utilize renewable energy and study the factors that favor renewable energy over traditional sources, especially for geographically distributed data centers \cite{hogade2018minimizing,liu2014greening,yuan2020revenue}. Most existing research efforts mainly focus on reducing computational energy and overlook cooling energy, a significant factor in terrestrial data centers \cite{wan2020sustainability} Moreover, renewable energy sources, such as solar and wind power, can be unreliable for mission-critical and large-scale data centers due to their dependence on variable weather conditions like cloud cover and wind patterns ~\cite{hu2019joint,agarwal2021redesigning}.

\subsection{Contributions}
To overcome the continual growth of data centers' energy consumption, non-traditional energy-efficient computation paradigms are needed. We advocate data center-enabled High Altitude Platform (HAP) as a practical and green alternative to terrestrial data centers \cite{abderrahim2023how}.   HAPs can be a core futuristic airborne network component that will revolutionize the networking frontier in the stratospheric range at an altitude between 17 km and 20 km \cite{abderrahim2023how, HAPEnabl,kurt2021vision}. HAPs offer several unique advantages, particularly from energy and ubiquity perspectives. Firstly, being located in the stratosphere saves cooling energy thanks to the naturally low atmospheric temperature, which ranges between $-50^{\circ}\text{C}$ and $15^{\circ}\text{C}$. Therefore, a HAP-enabled data center can offload some workloads from terrestrial data centers, saving the associated cooling energy. Additionally, HAPs can host large solar panels on their large surfaces that supply electricity to the data center servers and partially cover the required computational power \cite{abderrahim2023how, kurt2021vision, HAPEnabl}.
Secondly, HAPs have a pervasiveness advantage over terrestrial data centers thanks to the large footprint offered by the line of sight (LoS) links to the terrestrial infrastructure, wireless communication abilities and flexible relocation abilities \cite{abderrahim2023how, kurt2021vision, HAPEnabl}. HAPs can host multiple-input multiple-output (MIMO) antennas on their large surfaces and provide higher data rates to their users \cite{abderrahim2023how, kurt2021vision,ding2021joint}. However, it is necessary to define the limits and conditions under which HAP-enabled data centers can be beneficial, considering the practical limitations of HAPs, the quality of communication links, the range of proper offloaded workloads, and the impact of transmission delay on queuing time. Nonetheless, HAP-enabled data centers offer a promising and sustainable solution to the increasing energy consumption of traditional terrestrial data centers.

In this paper, we analyze the operation of a data center on a HAP considering  practical and realistic operational conditions. Firstly, we compare the energy models of HAP-enabled and terrestrial data centers, demonstrating significant energy savings with the former. We then analyze the harvested energy requirements to maintain the HAP's flying condition while performing the required computation for the offloaded terestrial workload. Next, we examine the reliability of the transmission link between terrestrial data centers and HAP-based systems by studying the outage probability. To address transmission outages, we propose a re-transmit dropped workloads solution using a portion of the saved energy. Finally, we explore the delay experienced in transmitting or re-transmitting workloads to a HAP-enabled data center without affecting the queuing time and verify the conditions that enable reliable operation. Our study presents an extensive and pioneering analysis of energy, communication, and delay performance aspects of HAP-enabled data centers. The system model is presented in Section 2, followed by our system analysis in Section 3. In Section 4, we discuss and analyze our results. We conclude the paper in Section 5.

\par Notations: Lower case boldface letters denote vectors while upper case boldface letters denote matrices.  The  conjugate-transpose of matrix $\boldsymbol{A}$ is denoted by $\boldsymbol{A}^H$ and the transpose of matrix $\boldsymbol{A}$ is denoted by $\boldsymbol{A}^T$ . $\norm{\cdot}_F^2$ denotes the square of Frobenius norm. $\norm{\cdot}$ denotes Euclidean norm. $\boldsymbol{I}_M$ denotes the ${M\times M}$ identity matrix. $\mathbb{C}^{M\times N}$ denotes the complex space of ${M\times N}$. $\text{Pr}(\cdot)$ denotes the probability. Moreover, Table \ref{table:1} summarizes the main notations used in this paper and their respective descriptions.

\newpage
 \begin{spacing}{1}
\begin{longtable}{|c|c |c|}
\caption{ Description of Main Notations}
\label{table:1}\\
 \hline
Notation &Description& Unit   \\ 
 \hline\hline
I & Number of servers in the terrestrial data center&\\
J & Number of cooling units in the terrestrial data center&\\
I' & Number of servers in the HAP&\\
$\lambda_i$ & Workload arrival rate to server $s_i$ & task/s\\
$u_i$ & Utilization ratio of server $s_i$&$\%$ \\
$\mu_i$ & Service rate of server $s_i$&MIPS\\
$\theta^*$ & Computational task length& Bits\\
$P_i^\text{idle}$  & Average power of idle server $s_i$ & Watt \\
$P_i^\text{peak}$ & Average power of fully utilized server $s_i$& Watt\\
$P^\text{fan}$ & Fan power& Watt\\
$Q_j$ & Heat amount removed by cooling unit ${ac}_j$&Watt\\
COP & Performance coefficient of cooling unit ${ac}_j$&$\%$\\
$P^\text{comp}_i$ & Computational power of server $s_i$& Watt\\
$E^\text{comp}_i$ & Computational energy of server $s_i$ & Joule\\
$E^\text{comp}_\text{TDC}$ & Computational energy of the terrestrial data center & Joule\\
$P^\text{cool}_j$ & Cooling power of cooling unit ${ac}_j$ & Watt\\
$E^\text{comp}_j$ &  Cooling energy of cooling unit ${ac}_j$ & Joule\\
$E^\text{cool}_\text{TDC}$ & Cooling energy of the terrestrial data center & Joule\\
l & HAP latitude & Degree\\
d & Considered day of the year&\\
$\eta_\mathrm{pv}$ & Efficiency of the photo-voltaic system of the HAP& $\%$\\
$\mathcal{A}_\mathrm{pv}$ & Area of the photo-voltaic system of the HAP & $\text{m}^2$\\
$G$ & Total extra-terrestrial solar radiance per $\text{m}^2$&$W/\text{m}^2$\\
$\rho$ & Air density&$kg/\text{m}^3$\\
$\eta_\text{prop}$ & Propeller efficiency&$\%$\\
$v_\text{wind}$ & Wind velocity& m/s\\
$v_\text{HAP}$ & HAP velocity& m/s\\
$C_\text{D}$ & Drag coefficient&$\%$ \\
$\Psi_0$ & Channel power gain at the reference distance&dB\\
$L_{\text{HAP}}$ & Distance between the terrestrial data center and HAP& m\\
$\zeta$ & Rician factor&\\
$f_\text{carrier}$ & Carrier frequency& Hz\\
$P^\text{harv}_\text{HAP}$ & Harvested power of the HAP & Watt\\
$E^\text{harv}_\text{HAP}$ & Harvested energy of the HAP&Joule\\
$P_\text{HAP}^\text{prop}$& Propulsion power of the HAP&Watt\\
$E_\text{HAP}^\text{prop}$& Propulsion energy of the HAP&Joule\\
$P_\text{HAP}^\text{payload}$ & Payload power of the HAP&Watt\\
$E_\text{HAP}^\text{payload}$ & Payload energy of the HAP&Joule\\
$E_\text{TDC-HAP}^\text{trans}$ & Transmission energy of the data center enabled-HAP&Joule\\
\hline
\end{longtable}
\end{spacing}

\section{System Model}
\begin{figure}
\centering
    \includegraphics[width=0.6\linewidth]{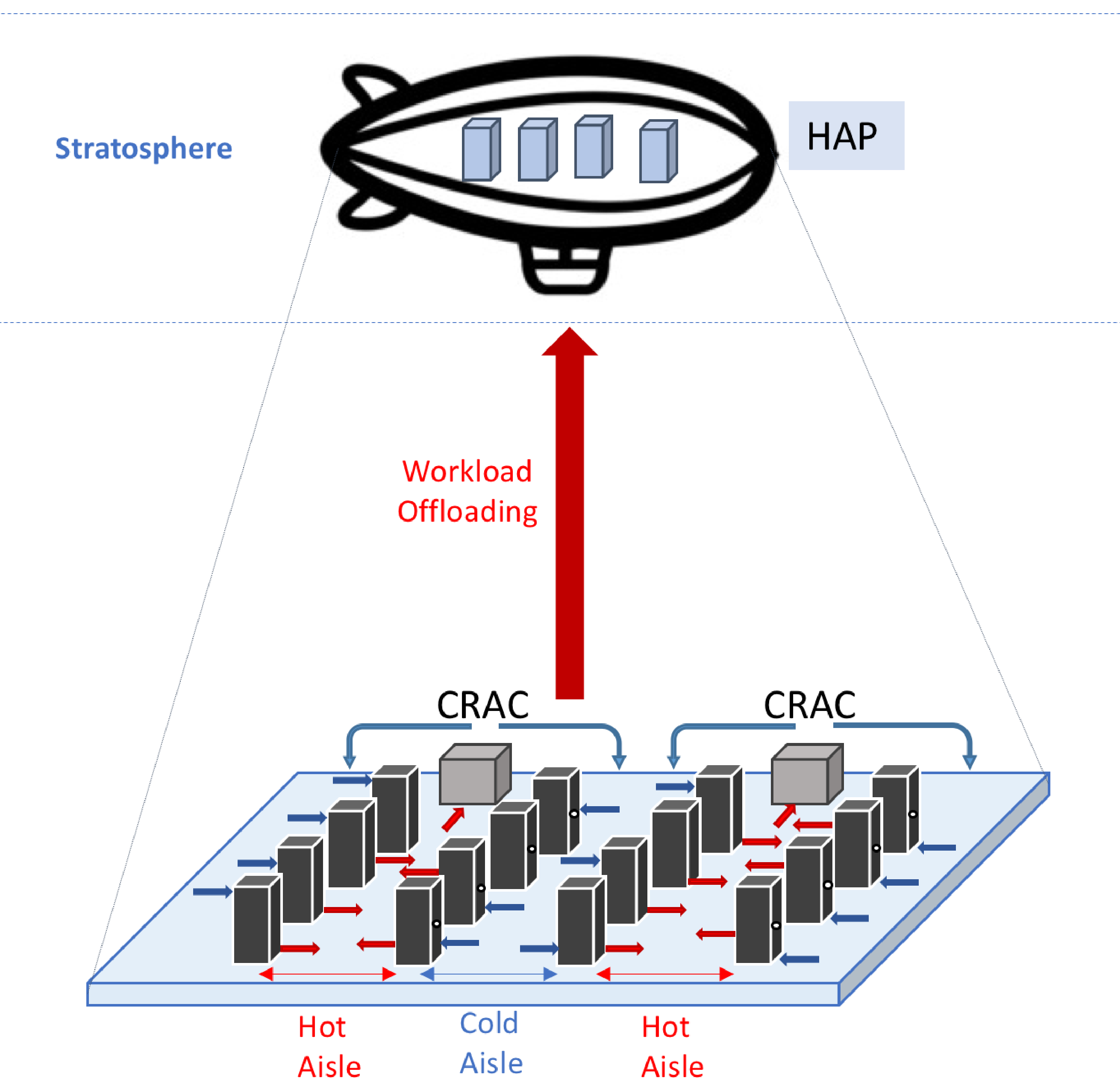}
    \caption{Data Center-enabled HAP Architecture}
    \label{fig:1}
\end{figure}

Our system model comprises a terrestrial data center and airborne data centers to offload some of the workloads. The terrestrial data center includes a set $\mathcal{I}={s_1,..., s_I}$ of $I$ servers arranged in a hot aisle/cold aisle configuration. The cold aisles face each other, while the hot aisles face each other. Additionally, the terrestrial data center has a cooling system consisting of a set $\mathcal{J}={{ac}_1 ,..., {ac}_J}$ of $J$ computer room air conditioning (CRAC) units, as shown in Fig.~\ref{fig:1}. The purpose of the CRAC units is to maintain a temperature range of $18^{\circ}\text{C}$ to $26^{\circ}\text{C}$, following the recommendation of the American Society of Heating, Refrigerating, and Air-Conditioning Engineers \cite{ashrae2021thermal}. On the other hand, the data center-enabled HAP comprises a set $\mathcal{I'}={s_{I+1},..., s{_I+I'}}$ of $I'<I$ servers. Since the stratosphere's average temperature falls between $-50^{\circ}\text{C}$ and $-15^{\circ}\text{C}$, we assume that the data center-enabled HAP does not require any CRAC units \cite{kurt2021vision, HAPEnabl}.
%Among the possible air ducting design for data-centers, we consider the "raised floor and ceiling return" schema \cite{li2017holistic}. 
%

\subsection{Workload Model}
%The workload arriving at our system between time $t_1$ and $t_2$ is a set $\mathcal{K}={\text{tk}_1,..., \text{tk}_K}$ of $K$ tasks. Each task $\text{tk}k$ has a length of $\theta_k$ instructions and requires a CPU utilization ratio of $u^\text{tk}k$. In the rest of this paper, we consider the task length and the task CPU utilization as random variables denoted by $\Theta^\text{tk}$ and $U^\text{tk}$, respectively. Based on this assumption, we further assume that all tasks have the same length $\theta^*=\mathbb{E}[\Theta^\text{tk}]= \sum{k=1}^K\left(\text{Pr}(\theta_k=\theta).:\theta\right)$ and the same CPU utilization $u^*=\mathbb{E}[U^\text{tk}]= \sum{k=1}^K\left(\text{Pr}(u^\text{tk}_k=u).:u\right)$. In other words, we assume that the expected values of the task length and the task CPU utilization are constant across all tasks.

% The writing seems technically accurate. However, I would suggest some minor changes to improve the clarity of the text. Here's a possible revision:

{The controller station (dispatcher) schedules the total workload that needs to be processed by our system among the servers $s_{i\in\{1,..., I+I'\}}$ in the terrestrial data center and data center-enabled HAP, according to a Poisson process \cite{li2017holistic}. Let $\Lambda$ denote the total workload arrival rate, such that $K=\int_{t_1}^{t_2}\Lambda(t)\:dt$, where $K$ is the total number of tasks arriving in the system. The number of tasks $K_i$ processed by server $s_i$ can be expressed as $K_i=\int_{t_1}^{t_2}\lambda_i(t)\:dt$, where $\lambda_i$ is the workload arrival rate to server $s_i$. We assume that the arrival rate follows a Poisson process when distributed by the controller station. Therefore, $\sum_{i=1}^{I+I'}{K_{i}}=\sum_{i=1}^{I+I'}\:\lambda_i\:(t_2-t_1)\leq K$.}

 {When the arriving workload is processed by server $s_i$, it transitions from the idle state to the active state and generates a CPU utilization ratio of $0<{u}_i(\lambda_i) \leq \overline{u}_i<100\%$, where $\overline{u}_i$ is the desirable utilization ratio of server $s_i$. We can express the utilization ratio ${u}_i$ as a function of the arrival rate ${u}_i(\lambda_i)= {\theta^*\:\lambda_i}/{\mu_i}$, where $\theta^*$ is the expected value of the task length assumed to be constant for all tasks. The service rate of server $s_i$, denoted by $\mu_i$ and measured in millions of instructions per second (MIPS), refers to the capacity of its CPU to process tasks.}

%Accordingly, the maximum workload rate $\lambda_i^\text{max}$ that can be scheduled to server $s_i$ is given by:
%\begin{align}
 %   \lambda_i^\text{max}= \dfrac{\mu_i\:\overline{u}_i}{\theta^*}\geq \lambda_i.
%\end{align}

%Moreover, each task requirement in terms of CPU usage should be guaranteed according to:

%\begin{equation}
 %  K_i\:u^*\leq \overline{u}_i.
%\end{equation}

\subsection{Power Model of the Terrestrial Data Center}
A breakdown of the energy consumed by a data center shows that the computational infrastructure and the cooling infrastructure are the two primary components that absorb the data center energy \cite{dayarathna2015}. Therefore, we focus on studying  the energy consumed by these components in the following subsections.
\subsubsection{Computational Model} 
The $I$ servers of the data center have different characteristics in terms of CPU capacity.  We assume %that the server utilization $u_i(\lambda_i(t))$ is equivalent to its CPU utilization and%
that the server power consumption and the CPU utilization have a linear relationship \cite{dayarathna2015}.   Therefore, the computational power $P^\text{comp}_i$ consumed by $s_i$ can be expressed as \cite{dayarathna2015,berezovskaya2020modular,zhabelova2018comprehensive}:
\begin{equation}
P^\text{comp}_i(\lambda_i)=P_i^\text{idle}+\left(P_i^\text{peak}-P_i^\text{idle}\right)\:u_i(\lambda_i),
\label{eq:comp1}
\end{equation} 
where $P_i^\text{idle}$  is the average power value when  server $s_i$ is idle (i.e. $u_i(\lambda_i(t))=0\%$) and $P_i^\text{peak}$ is the average power value when the server is fully utilized, i.e., $u_i(\lambda_i(t))=\overline{u}_i$. 

\subsubsection{Cooling Model} 
The compressor and the fans of the CRAC units are the main consumers of the cooling energy in a data center \cite{li2017holistic,dayarathna2015}.  In a data center, the cold air supplied by the CRAC units enters through the cold aisle to the server inlets in the front of the server racks and is exhausted into the hot aisle through the server outlets in the back of the server racks. As a result of this air circulation, each node in the data center, whether a server $s_{i\in{1..I}}$ or a CRAC unit ${ac}_{j\in{1..J}}$, has an inlet temperature $T^\text{in}$ and an outlet temperature $T^\text{out}$. The inlet temperature $T^\text{in}$ represents the amount of heat received from other nodes, while the outlet temperature $T^\text{out}$ represents the amount of heat contained in a given node. Therefore, we can express the cooling power $P^\text{cool}j$ of CRAC unit ${ac}_{j\in{1..J}}$ as the power necessary to cool the servers under its coverage. This cooling power can be calculated as follows \cite{li2017holistic,parolini2011cyber},
\begin{equation}
    P^\text{cool}_j(t)=P_j^\text{fan}+ \dfrac{Q_j(t)}{\text{COP}(T_{j}^\text{out})},
    \label{eq:Pcool}
\end{equation}
where $P^\text{fan}$ is the fan power, {$Q_j$ is the heat amount removed by CRAC unit ${ac}_j$}  and COP is the ${ac}_j$ performance coefficient given by $\text{COP}(T_{j}^\text{out})=0.0068\:\left(T_{j}^\text{out}\right)^2+0.008\:T_{j}^\text{out}+0.458$.

%The fans $f_{k\in\{1..K\}}$ present in a data center are another main consumer of energy \cite{dayarathna2015}. The  cooling power $P^\text{fan}_k$ of  fan $f_{k\in\{1..K\}}$   increases linearly to the cube of the fan speed $w_k$ and can be expressed as \cite{dayarathna2015,chan2018optimal}:
%\begin{equation}
%P^\text{fan}_k(d)= \frac{P^\text{fan}_\text{ref}}{w_\text{ref}^3}\:w_k^3(d);
%\end{equation}%
%where $P^\text{fan}_\text{ref}$ is the consumed power for a known fan speed $w_\text{ref}$. 

\subsection{Power Model of the Data Center-enabled HAP}
Managing energy in HAPs is crucial because they are typically designed for long-duration missions (at least one year). Throughout this subsection, we study the energy harvested and consumed by a data center-enabled HAP.
\subsubsection{Energy Harvesting Model} 
 The main energy source for the daytime operation of HAPs is the solar energy  \cite{kurt2021vision,HAPEnabl}. HAPs also incorporate energy storage components; which are typically Lithium-Sulphur batteries or hydrogen fuel cells \cite{arum2020energy,kurt2021vision}.  These batteries support the nighttime operation of the HAP and are fed by the solar energy harvested during the daytime \cite{arum2020energy,kurt2021vision}. By assuming that the energy harvested can be stored for $24\:h$,  the  mean solar power harvested  by the HAP at latitude $l$ on day $d$ can be expressed as \cite{arum2020energy},
\begin{equation}
    P^\text{harv}_\text{HAP}(l,d)=\eta_\mathrm{pv}\:\mathcal{A}_\mathrm{pv}\:G(l,d),
\end{equation}
where ${\eta}_\mathrm{pv}$ is the efficiency of the photo-voltaic system,  $\mathcal{A}_\mathrm{pv}$ is the area of the photo-voltaic system and $G$ is the total extra-terrestrial solar radiance per $\text{m}^2$ and is given by \cite{arum2020energy}, 
%$G$ refers to the solar radiation detected above the atmosphere 
\begin{equation}
    G(l,d)= \dfrac{ \tau(l,d) G_\text{max}(l,d)}{\xi_\text{max}(l,d)} \left(1-\cos(\xi_\text{max}(l,d))\right),
\end{equation}
where $\tau(l,d)$ is the daylight duration at latitude $l$ on day $d$,  $G_\text{max}(l,d)$ is corresponding  maximum radiation intensity and $\xi_\text{max}(l,d)$ is the corresponding maximum altitude of the sun (in rd). $\tau(l,d)$ is expressed in hours from \cite{arum2020energy},
\begin{equation}
    \tau(l,d)=  1-\frac{1}{\pi} \cos^{-1}\left( \dfrac{\tan(l)\sin(\epsilon)\sin(\phi(d))}{\sqrt{1-\sin^2(\epsilon)\sin^2(\phi(d))}} \right),
\end{equation}
where $\epsilon=0.4093$ is the angle of obliquity of the Earth and $\phi(d)$ is the azimuthal angle of the sun and is given by \cite{arum2020energy},
\begin{align}
    \phi(d)=& -1.3411+M(d)+0.0334 \sin(M(d))  + 0.0003\sin(2M(d)),
\end{align}
such that $M(d)=-0.041+0.017202d$ is the angular distance of the sun, which is also known as mean anomaly \cite{arum2020energy}.  
 $G_\text{max}(l,d)$ can be computed from \cite{arum2020energy},
\begin{equation}
\label{eq:rad_int}
    G_\text{max}(l,d)=G_\text{solar} \left( 1+0.033 \cos\left(\frac{360 d}{365}\right) \right)\cos\left(l-\delta(d)\right),
\end{equation} 
where $G_\text{solar}=1366.1\:{W/m^2}$ is the standard solar constant at zero air mass and $\delta$  is the solar declination angle expressed as (in rd) \cite{arum2020energy}:
\begin{equation}
    \delta(d)=0.4093\:\sin\left(\dfrac{2\pi\:(d-79.75)}{365}\right).
\end{equation}
Whereas $\xi_\text{max}$ is given by  $\xi_\text{max}(l_r,d)=\frac{\pi}{2}+l_r-\delta(d)$;  where $l_{\mathrm{r}}$ is the latitude in rd \cite{arum2020energy}.

\subsubsection{Consumption Model} 
The power consumed by a HAP is determined mainly by  both the payload and the propulsion subsystems \cite{arum2020energy,kurt2021vision}. The required power by the payload is the computational power of the servers carried by the HAP. The required power by the HAP propeller can be expressed as \cite{sun2020simulation},
 \begin{equation}
     P_\text{HAP}^\text{prop}(l,d)=\:\dfrac{\rho}{2\eta_\text{prop}}\:v_\text{wind}^3(l,d)\: v_\text{HAP}^{2/3}\:C_\text{D},
     \label{eq:Pprop}
 \end{equation}
  where $\rho$ is the air density, $\eta_\text{prop}$ is the propeller efficiency, $v_\text{wind}$ is the wind velocity, $v_\text{HAP}$ is the HAP velocity and $C_\text{D}$ is the drag coefficient, which is expressed as $C_D=N_\mathrm{C} \:C_\text{envelope}$; such that $N_\mathrm{C}$ is a constant equal to $1.8$ in the scope of this paper and $C_\text{envelope}$ is given by  \cite{sun2020simulation},
  \begin{equation}
    C_\text{envelope}  = \dfrac{0.172 \:f_\mathrm{r}^{\frac{1}{3}}+0.252\: f_\mathrm{r}^{-1.2}+1.032\:f_\mathrm{r}^{-2.7}}{R_\mathrm{e}^{\frac{1}{6}}},
  \end{equation}
  where $f_\mathrm{r}=\frac{L_\text{HAP}}{D_\text{HAP}}$ is the fitness ratio of the HAP body length $L_\text{HAP}$ to its maximum width/diameter $D_\text{HAP}$, and $R_\mathrm{e}$ is the Reynolds number given by \cite{sun2020simulation},
  \begin{equation}
  R_\mathrm{e}= \dfrac{\rho\: v_\text{wind}(l,d)\: D_\text{HAP}}{\kappa},
  \end{equation}
  where $\kappa$ is the air dynamic viscosity \cite{sun2020simulation}.

\subsection{Channel Model}
We assume that both the terrestrial data center and the data center-enabled HAP utilize MIMO technology to overcome path loss and fading \cite{ding2021joint}. Specifically, the terrestrial data center is equipped with $N$ antennas, and the HAP is equipped with $M$ antennas. The channel between the terrestrial data center and the HAP can be modeled using Rician fading, which accounts for both a line-of-sight component and a scattered component. This channel model can be expressed as follows \cite{ding2021joint,lian2019user,ji2020energy,michailidis2010three}:
\begin{equation}
    \boldsymbol{\boldsymbol{H}}_{\text{HAP}}(l)=\sqrt{\dfrac{\Psi_0}{L^2_{\text{HAP}}}}\left(\sqrt{\frac{\zeta}{\zeta+1}}\bar{\boldsymbol{H}}_{\text{HAP}}(l)+\sqrt{\dfrac{1}{\zeta+1}}\Hat{\boldsymbol{H}}_{\text{HAP}}\right), %\in \mathbb{C}^{M\times N};
\end{equation}
where $\Psi_0$ is the channel power gain at the reference
distance; $L_{\text{HAP}}$ is the distance between the terrestrial data center and the  HAP; $\zeta$ is the Rician factor;  $\bar{\boldsymbol{H}}_{\text{HAP}} \in \mathbb{C}^{M\times N}$ is the LoS channel component and $\Hat{\boldsymbol{H}}_{\text{HAP}} \in \mathbb{C}^{M\times N}$ is the Rayleigh fading component. The LoS channel component is given by $\bar{\boldsymbol{H}}_{\text{HAP}}(l)=\exp\left({-j \frac{2\pi f_\text{carrier}}{C}\: r_{\text{HAP}}}(l)\right)$ where $f_\text{carrier}$ is the carrier frequency; $C$ is the speed of light and $r_{\text{HAP}}$ is the length of the direct path between the transmit and receive antennas. The Rayleigh fading component $\Hat{\boldsymbol{H}}_{\text{HAP}}$ follows the distribution $\mathcal{CN}(0,\boldsymbol{I})$. Accordingly, the achievable data rate between the terrestrial data center and the HAP can be expressed as \cite{ding2021joint},
\begin{equation}
\footnotesize
    R_{\text{HAP}}(l)=B \log\det\left(\boldsymbol{I}_M+\boldsymbol{H}_{\text{HAP}}(l) \boldsymbol{q}_{\text{HAP}}\boldsymbol{q}_{\text{HAP}}^H \boldsymbol{H}_{\text{HAP}}^H(l) \boldsymbol{n}^{{-1}}\right), \normalsize	
\end{equation}
 where $B$ is the bandwidth available for the transmission between the terrestrial data center and the HAP; $\boldsymbol{\boldsymbol{q}}_{\text{HAP}} \in \mathbb{C}^{N\times {1}}$ is the precoding matrix of the terrestrial data center; $\boldsymbol{n} \in \mathbb{C}^{M\times1}$ is the additive white Gaussian noise (AWGN) with distribution $\mathcal{CN}(0,\sigma^2\boldsymbol{I}_M)$.

\section{System Analysis}
In this section, we analyze the performance of a data center-enabled HAP and compare it with a terrestrial data center. We begin by examining the conditions under which the energy harvested by the HAP's solar panels can sustain the HAP's flight and power the hosted servers. Next, we establish a comprehensive energy model for both systems and demonstrate that a data center-enabled HAP can achieve significant energy savings compared to a terrestrial data center. We then investigate the outage probability of the transmission link between the terrestrial data center and the HAP and explore whether the saved energy can be utilized to manage dropped workloads. Finally, we analyze the delay experienced in a data center-enabled HAP, which consists of the transmission delay and the mean waiting time that represents the queuing time of tasks in the servers before processing.
To facilitate our analysis, we define the workload arrival rate vector for the terrestrial data center as $\boldsymbol{\lambda}_\text{TDC}={[\lambda_{i}]}\in \mathbb{R}^{I}$ and the workload arrival rate vector for the data center-enabled HAP as $\boldsymbol{\lambda}_\text{HAP}={[\lambda_{i}]}\in \mathbb{R}^{I'}$. By using these vectors, we can compare and evaluate the workload performance of the two systems.

\subsection{HAP Flying Condition}
Throughout the following analysis,  we aim to determine whether the energy harvested by the solar panels on the HAP ($E_\text{HAP}^\text{harv}$) is sufficient to meet the energy requirements of both the payload ($E_\text{HAP}^\text{payload}$) and propulsion ($E_\text{HAP}^\text{prop}$). In our study, we consider the HAP payload to be a modular data center comprising a set of $I'$ servers denoted by $\mathcal{I'}={s_{I+1},..s_{I+I'}}$. The power consumption of the payload is determined by the computational power of the servers, as given in equation (\ref{eq:comp1}). Therefore, the total power consumption of the payload can be expressed as,
  \begin{align}
      P_\text{HAP}^\text{payload}(\boldsymbol{\lambda}_\text{HAP})=%\sum_{i=I+1}^{I+I'} \left(P^\text{comp}_i(\lambda_i)\right)\\
      \sum_{i=1}^{I'}\left(P_i^\text{idle}+\left(P_i^\text{peak}-P_i^\text{idle}\right) 
      u_i(\lambda_i)\right).
      \label{eq:Ppay}
  \end{align}
  Hence, $E_\text{HAP}^\text{payload}$ can be determined by integrating equation (\ref{eq:Ppay}), resulting in the following expression:
  \begin{align}
  \label{eq:Epay}
      E_\text{HAP}^\text{payload}(\boldsymbol{\lambda}_\text{HAP})%&= \int_{t_1}^{t_2} P_\text{HAP}^\text{payload}(t)\: dt\\
      \!=\!\!\!\!\sum_{i=I+1}^{I+I'}\!\!\left(\!P_i^\text{idle}+\!\dfrac{\lambda_i\theta^*}{\mu_i}\!\left(P_i^\text{peak}\!-\!P_i^\text{idle}\!\right)\!\right)\!(t_2-t_1).
  \end{align}
Similarly, $E_\text{HAP}^\text{prop}$ is derived by integrating equation (\ref{eq:Pprop}), which yields
\begin{equation}
\label{eq:Eprop}
 E_\text{HAP}^\text{prop}(l,d)=v_\text{wind}^{\frac{17}{6}}(l,d)\:C^{'}_D\:(t_2-t_1),  
\end{equation}
where\begin{equation} C^{'}_D \!= \!\dfrac{\rho^\frac{5}{6}  v_\text{HAP}^{2/3}}{2\eta_\text{prop}}  \dfrac{\kappa^{\frac{1}{6}} N_\mathrm{C} \! \left(0.172  f_\mathrm{r}^{\frac{1}{3}}+0.252  f_\mathrm{r}^{-1.2}\!+1.032 f_\mathrm{r}^{-2.7}\right)}{D_\text{HAP}^{\frac{1}{6}}}.\end{equation}

Therefore, we define the flying condition of the HAP as follows:
\begin{equation}
    E_\text{HAP}^\text{prop}(l,d)+E_\text{HAP}^\text{payload}(\boldsymbol{\lambda}_\text{HAP}) \leq E^\text{harv}_\text{HAP}(l,d).
    \label{eq:flyCond}
\end{equation}

The flying condition of the HAP is affected by three parameters:  $l$, $d$ and  $\boldsymbol{\lambda}_\text{HAP}$. To maintain the desired flying condition, we control the workload arrival rate that can be processed by the servers on the HAP, i.e., $\boldsymbol{\lambda}_\text{HAP} \in \mathbb{R}^{I'}$. Specifically, for a fixed latitude $l^*$ and day of operation $d^*$, the workload arrival rate $\boldsymbol{\lambda}_\text{HAP}$ must satisfy the constraint $\boldsymbol{\lambda}_\text{HAP} \leq \boldsymbol{\lambda}_\text{HAP}^\text{max}$, where $\boldsymbol{\lambda}_\text{HAP}^\text{max}$ is the maximum allowable workload arrival rate. Hence, the corresponding energy consumption is 
%\begin{equation}
  $ E_\text{HAP}^\text{payload}(\boldsymbol{\lambda}_\text{HAP}^\text{max})= E^\text{harv}_\text{HAP}(l^*,d^*)- E_\text{HAP}^\text{prop}(l^*,d^*).$
  %  \label{eq:flyCond}
%\end{equation}

Assuming symmetrical characteristics for the servers, including the same service rate ($\mu_i=\mu$), desirable utilization ratio ($\overline{u}i=\overline{u}$), and power consumption ($P_i^\text{idle}=P^\text{idle}$ and $P_i^\text{peak}=P^\text{peak}$), the workload can be evenly distributed across multiple servers with $\lambda{i}^\text{max}=\lambda^\text{max}$. To determine the maximum allowable workload arrival rate $\lambda^\text{max}$, we can use the following expression:
 \begin{align}
\label{eq:lambda_max}
 &\lambda^\text{max}(l^*,d^*)\!= \Big(\dfrac{\eta_{pv}\mathcal{A}_{pv}  \tau(l^*,d^*) G_\text{max}(l^*,d^*)\left(1\!-\!\cos(\xi_\text{max}(l^*,d^*))\right)}{24 \xi_\text{max}(l^*,d^*)}\nonumber\\ &-v_\text{wind}^{\frac{17}{6}}(l^*,d^*)C^{'}_D-I'P^\text{idle}\Big)\dfrac{\mu}{I'\theta^*\left(P^\text{peak}-P^\text{idle}\right)}. 
\end{align}

\subsection{Energy Saving}
 In this section, we investigate and compare the energy consumed by a terrestrial data center $E_\text{TDC}^\text{cons}$ to the energy consumed by a data center-enabled HAP $E_\text{TDC-HAP}^\text{cons}$. Hence, we assume that all the workload directed to the HAP is successfully transmitted and no link outage is experienced. To establish a fair comparison, we assume that both systems have $I+I'$ servers in total. Therefore, $I+I'$ servers are placed in the terrestrial data center. In the data center-enabled HAP, $I$ servers are placed in the terrestrial data center and $I'$ servers are placed in the HAP.
 \subsubsection{ Energy Consumption in Terrestrial Data Center } 
 To understand the energy consumption of a terrestrial data center, i.e.,  $E_\text{TDC}^\text{cons}$,  we break it down into two components: the computational energy consumed by the servers ($E^\text{comp}_\text{TDC}(\boldsymbol{\lambda}_\text{TDC})$) and the cooling energy consumed by CRAC units ($E^\text{cool}_\text{TDC}(\boldsymbol{\lambda}_\text{TDC})$), which is written as, 
 $$E_\text{TDC}^\text{cons}(\boldsymbol{\lambda}_\text{TDC}) = E^\text{comp}_\text{TDC}(\boldsymbol{\lambda}_\text{TDC}) + E^\text{cool}_\text{TDC}(\boldsymbol{\lambda}_\text{TDC}).$$
 $E^\text{comp}_\text{TDC}$ captures the computational energy consumed by all the servers expressed as,
$$E^\text{comp}_\text{TDC}(\boldsymbol{\lambda}_\text{TDC})=\sum_{i=1}^{I+I'} E^\text{comp}_i.$$
To calculate the computational energy consumed by each server, we integrate equation (\ref{eq:comp1}) and obtain the value of $E^\text{comp}_i$ as,
\begin{align}
    E^\text{comp}_i(\lambda_i)=\left(P^\text{idle}+\dfrac{\lambda_i\:\theta^*}{\mu}\left(P^\text{peak}-P^\text{idle}\right)\:\right)(t_2-t_1).
    \label{eq:Ecomp}
\end{align}

 $E^\text{cool}_\text{TDC}$ captures the energy consumed by the $J$ CRAC units, i.e., $$E^\text{cool}_\text{TDC}(\boldsymbol{\lambda}_\text{TDC})=\sum_{j=1}^J E^\text{cool}_j,$$ where the cooling energy consumed by CRAC unit $j$, denoted as  $E^\text{cool}_j$, is obtained by integrating (\ref{eq:Pcool}) as,
\begin{equation}
      E^\text{cool}_j= \int_{t_1}^{t_2}P_j^\text{fan}+\dfrac{Q_j(t)}{\text{COP}(T_{j}^\text{out})}\:dt,
      \label{eq:Ecool}
\end{equation}
where $P_j^\text{fan}$ is computed from 
$$P_j^\text{fan}=\dfrac{\Dot{V}\Delta p^\text{fan}}{\eta_j^\text{fan}\eta_j^\text{fan,motor}},$$ with $\Dot{V}$ denoting the air flow rate of the CRAC unit, $\Delta p^\text{fan}$ is the pressure loss due to air flow resistances, $\eta_j^\text{fan}$ and $\eta_j^\text{fan,motor}$ are the efficiencies of the fan and motor fan, respectively \cite{erden2017experimental}.

The heat removed by CRAC unit ${ac}_j$ from $I_1<I$ servers in the data center, i.e., $Q_j(t)$  is found from \cite{berezovskaya2020modular},
\begin{equation}
Q_j(t)=\sum_{i=1}^{I_1} \left(C_\text{air} \:f_\text{air}\: (T^\text{out}_i(t)-T^\text{in}_i(t))\right),
\label{eq:heat}
\end{equation}
where $C_\text{air}$ is the air heat capacity, $f_\text{air}$ is the air flow rate through the server CPU; $T^\text{in}_i$ and $T^\text{out}_i$ are respectively the inlet temperature and  the outlet temperature of server $s_i$. $T^\text{in}_i$ is expressed as \cite{li2017holistic},
\begin{equation}
    T^\text{in}_i(t)=T_j^\text{out}+\gamma+\big(T^\text{in}_i(0)-T_j^\text{out}-\gamma\big)e^{-\nu\:t},
\label{Tin}
\end{equation}
where $T^\text{in}_i(0)$ is the initial inlet temperature of server $s_i$; $\gamma$ is the temperature raise imposed by the recirculated exhausted air in the data center and $\nu$ reflects the temperature influence of the closest CRAC. On the other handm, $T^\text{out}_i$ is found from  \cite{berezovskaya2020modular},
\begin{equation}
T^\text{out}_i(t)=\left(1-\frac{1}{C_\text{air} f_\text{air}R}\right)T^\text{in}_i(t)+\frac{1}{C_\text{air} f_\text{air}R} T^\text{CPU}_i(t), \label{eq:outlet}
\end{equation}
where $R$ is the thermal resistance of the CPU and $T^\text{CPU}_i$ is the CPU temperature such that \cite{li2017holistic,WOLF2017149}:
\begin{equation}
    T^\text{CPU}_i\!(t)\!\!=\!\!T^\text{in}_i(t)+RP^\text{comp}_i+\!\bigg( \! T^\text{CPU}_i\!(0)-T^\text{in}_i\!(t)-RP^\text{comp}_i \!\! \bigg)e^{\!-\!\frac{t}{RC_i}},
\label{TCPU}
\end{equation}
where $T^\text{CPU}_i(0)$ is the CPU temperature at the initial time and $C_i$ is the server heat capacity. 
Substituting (\ref{Tin}),  (\ref{eq:outlet}) and (\ref{TCPU})  in (\ref{eq:heat}) yields: 
\begin{equation}
\begin{split}
    Q_j(t)=&\frac{1}{R}\sum_{i=1}^{I_1}\bigg(\!\!R(1-e^{-\frac{t}{RC_i}}) P^\text{comp}_i\!+\!(T^\text{CPU}_i(0)\!-\!T_j^\text{out}-\gamma)e^{-\frac{t}{R C_i}}
    -\big(T^\text{in}_i(0)-T_j^\text{out}-\gamma\big) e^{-(\nu+\frac{1}{R C_i})t}\bigg).
    \label{eq:Qj}
\end{split}
\end{equation}
By substituting (\ref{eq:Qj}) in (\ref{eq:Ecool}), $E^\text{cool}_j$ consumed by ${ac}_j$ can be re-written as,
\begin{align}
    &E^\text{cool}_j=  \sum_{i=1}^{I_1}\dfrac{C_i(T^\text{CPU}_i(0)-T_j^\text{out}-\gamma)(e^{-\frac{t_1}{RC_i}}-e^{-\frac{t_2}{RC_i}})}{\left(0.0068\left(T_{j}^\text{out}\right)^2+0.008T_{j}^\text{out}+0.458\right)}\nonumber\\
    &+\sum_{i=1}^{I_1}\dfrac{C_i(T^\text{in}_i(0)-T_j^\text{out}-\gamma)(e^{-(\nu+\frac{1}{R\:C_i})t_2}-e^{-(\nu+\frac{1}{RC_i})t_1})}{(\nu R C_i+1) \left(0.0068 \left(T_{j}^\text{out}\right)^2+0.008 T_{j}^\text{out}+0.458\right)}\nonumber\\
    &+\sum_{i=1}^{I_1} \!\dfrac{\left(\!P^\text{idle}\!+\!\dfrac{\lambda_i \theta^*}{\mu}\!\left(\!P^\text{peak}\!-\!P^\text{idle}\!\right) \!\right)\!\left(\!t_2\!-\!t_1 \!\right)}{0.0068\! \left(T_{j}^\text{out}\right)^2\!+\!0.008 T_{j}^\text{out}\!+\!0.458}\!+\!P_j^\text{fan}\!\left(\!t_2\!-\!t_1\!\right)\nonumber\\
    &+\sum_{i=1}^{I_1} \dfrac{\left(\!P^\text{idle}\!+\!\dfrac{\lambda_i \theta^*}{\mu}\!\left(\!P^\text{peak}\!-\!P^\text{idle}\right)\! \right)\!\left(\!RC_i \!\left(\!e^{-\frac{t_2}{R C_i}}\!-\!e^{-\frac{t_1}{C_i}}\!\right)\!\right)}{0.0068 \left(T_{j}^\text{out}\right)^2+0.008 T_{j}^\text{out}+0.458}.
    \label{eq:Ecool2}
\end{align}  By summing over (\ref{eq:Ecomp})  and (\ref{eq:Ecool2}),  $E_\text{TDC}^\text{cons}(\boldsymbol{\lambda}_\text{TDC})$ can be re-written as,
 \begin{align}
 \label{eq:consTDC}
&E_\text{TDC}^\text{cons}(\boldsymbol{\lambda}_\text{TDC})=\!\sum_{i=1}^{I+I'}\!\left(\!P^\text{idle}+\dfrac{\lambda_i \theta^*}{\mu}\left(P^\text{peak}-P^\text{idle}\!\right) \!\right)\!(t_2-t_1)+\sum_{j=1}^J P_j^\text{fan}\left(t_2-t_1\right)\nonumber\\
&+\sum_{i=1}^{\frac{I+I'}{J}} \dfrac{C_i\:(T^\text{CPU}_i(0)-T_j^\text{out}-\gamma) (e^{-\frac{t_1}{R C_i}}-e^{-\frac{t_2}{R C_i}})}{\left(0.0068 \left(T_{j}^\text{out}\right)^2+0.008 T_{j}^\text{out}+0.458\right)}\nonumber\\
&+\sum_{i=1}^{\frac{I+I'}{J}} \dfrac{C_i (T^\text{in}_i(0)-T_j^\text{out}-\gamma)\:(e^{-(\nu+\frac{1}{R C_i})t_2}-e^{-(\nu+\frac{1}{R C_i})t_1})}{(\nu R C_i+1) \left(0.0068 \left(T_{j}^\text{out}\right)^2+0.008 T_{j}^\text{out}+0.458\right)}\nonumber\\
&+\sum_{i=1}^{\frac{I+I'}{J}} \dfrac{\left(P^\text{idle}+\dfrac{\lambda_i \theta^*}{\mu}\left(P^\text{peak}-P^\text{idle}\right) \right)\left(t_2-t_1\right)}{0.0068 \left(T_{j}^\text{out}\right)^2+0.008 :T_{j}^\text{out}+0.458}\nonumber\\
&+\sum_{i=1}^{\frac{I+I'}{J}} \dfrac{\left(\!P^\text{idle}\!+\!\dfrac{\lambda_i\theta^*}{\mu}\!\left(\!P^\text{peak}\!-\!P^\text{idle}\right) \! \right)\!R C_i \!\left( \! e^{-\frac{t_2}{R C_i}} \! -e^{-\frac{t_1}{R C_i}} \! \right)}{0.0068 \left(T_{j}^\text{out}\right)^2+0.008 T_{j}^\text{out}+0.458}.
\end{align}

\subsubsection{ Energy Consumption in  Data Center-enabled HAP} 
The energy consumed by the data center-enabled HAP is expressed as a function of  $l$, $d$ and  the  processed workload $\boldsymbol{\lambda}= [\boldsymbol{\lambda}_\text{HAP}^T \;\boldsymbol{\lambda}_\text{TDC}^T]^T \in \mathbb{R}^{I+I'} $:
 \begin{align}
    & E_\text{TDC-HAP}^\text{cons}(\boldsymbol{\lambda},l,d)=E_\text{HAP}^\text{payload}(\boldsymbol{\lambda}_\text{HAP})+E_\text{HAP}^\text{prop}(l,d)\nonumber\\
    &+E_\text{TDC-HAP}^\text{trans}(l,\boldsymbol{\lambda}_\text{HAP})+E^\text{comp}_\text{TDC-HAP}(\boldsymbol{\lambda}_\text{TDC})+E^\text{cool}_\text{TDC-HAP}(\boldsymbol{\lambda}_\text{TDC}).
         \label{eq:consHT}
    \end{align}
 Specifically, $E_\text{TDC-HAP}^\text{cons}$ comprises the computational energy (\ref{eq:Ecomp}) and the cooling energy  (\ref{eq:Ecool2}) besides  the payload energy (\ref{eq:Epay}), the propulsion energy of the HAP  (\ref{eq:Eprop}) and the transmission energy.  The transmission energy required to send the workload from the terrestrial data center to the HAP is expressed as \cite{ding2021joint}:
\begin{equation}
\label{eq:E_trans}
    E_\text{TDC-HAP}^\text{trans}(l,\boldsymbol{\lambda}_\text{HAP})=\dfrac{\beta\:b(\boldsymbol{\lambda}_\text{HAP})}{R_{\text{HAP}}(l)}\norm{\boldsymbol{q}_{\text{HAP}}}_F^2\:(t_2-t_1),
\end{equation}
where $\beta$ is the ratio of the transmitted data size to the original task data size due to the transmission overhead; $b$ is the size of the input data (in bits)   and $\norm{\boldsymbol{q}_{\text{HAP}} }_F^2$ is the transmit power from the terrestrial data center to the HAP.
The size of the input data $b$ can be expanded as $b(\boldsymbol{\lambda}_\text{HAP})=\sum_{i=I+1}^{I+I'}\left(\lambda_{i}\:\theta^*\:b^*\right)$ such that $b^*$ is the size of the instruction in bits. 

    By substituting   (\ref{eq:Epay}), (\ref{eq:Eprop}), (\ref{eq:E_trans}) in (\ref{eq:consHT}) and summing over (\ref{eq:Ecomp}) and (\ref{eq:Ecool2}),  the energy consumed by the data center-enabled HAP is re-written as:
    \begin{align}
     & E_\text{TDC-HAP}^\text{cons}(\boldsymbol{\lambda},l,d)= v_\text{wind}^{\frac{17}{6}}(l,d) C^{'}_D (t_2-t_1)+\dfrac{\beta b}{R_{\text{HAP}}(l)}\norm{\boldsymbol{q}_{\text{HAP}}}_F^2 (t_2-t_1)+\sum_{j=1}^J P_j^\text{fan}\left(t_2-t_1\right)\nonumber\\
     &+\sum_{i=1}^{I+I'}\left(P^\text{idle}+\dfrac{\lambda_i \theta^*}{\mu}\left(P^\text{peak}-P^\text{idle}\right) \right)(t_2-t_1)+\sum_{i=1}^{\frac{I}{J}} \dfrac{C_i (T^\text{CPU}_i(0)-T_j^\text{out}-\gamma) (e^{-\frac{t_1}{R C_i}}-e^{-\frac{t_2}{R C_i}})}{\left(0.0068 \left(T_{j}^\text{out}\right)^2+0.008 T_{j}^\text{out}+0.458\right)}\nonumber\\
&+\sum_{i=1}^{\frac{I}{J}} \dfrac{C_i (T^\text{in}_i(0)-T_j^\text{out}-\gamma) (e^{-(\nu+\frac{1}{R\:C_i})t_2}-e^{-(\nu+\frac{1}{R C_i})t_1})}{(\nu R C_i+1) \left(0.0068 \left(T_{j}^\text{out}\right)^2+0.008 T_{j}^\text{out}+0.458\right)}\nonumber\\
&+\sum_{i=1}^{\frac{I}{J}} \dfrac{\left(P^\text{idle}+\dfrac{\lambda_i \theta^*}{\mu}\left(P^\text{peak}-P^\text{idle}\right) \right)\left(t_2-t_1\right)}{0.0068 \left(T_{j}^\text{out}\right)^2+0.008 T_{j}^\text{out}+0.458}\nonumber\\
&+\sum_{i=1}^{\frac{I}{J}} \dfrac{\!\left(P^\text{idle}\!+\!\dfrac{\lambda_i\theta^*}{\mu}\!\left(\!P^\text{peak}\!-\!P^\text{idle}\!\right) \!\right)\!R C_i \!\left(\!e^{-\frac{t_2}{R C_i}}\!-\!e^{-\frac{t_1}{R C_i}}\!\right)}{0.0068 \left(T_{j}^\text{out}\right)^2+0.008 T_{j}^\text{out}+0.458}.
 \end{align}

 We propose that using a data center-enabled high-altitude platform (HAP) is a more energy-efficient option compared to a terrestrial data center. This is because the HAP does not require cooling units, as the average temperature in the stratosphere is significantly lower than the recommended temperature for a data center. Additionally, the HAP utilizes solar energy harvested during the daytime and stored in Lithium-Sulphur batteries during the nighttime to power its servers, whereas a terrestrial data center relies on electric energy supplied through the electrical grid constantly. As a result, the energy saved by using the HAP can be expressed as the difference between the energy consumed by the HAP data center, denoted as $E_\text{TDC-HAP}^\text{cons}(\boldsymbol{\lambda},l,d)$, and that consumed by the terrestrial data center, denoted as $E_\text{TDC}^\text{cons}(\boldsymbol{\lambda}\text{TDC})$, which can be written as $E^\text{sav}(\boldsymbol{\lambda},l,d)=E\text{TDC-HAP}^\text{cons}(\boldsymbol{\lambda},l,d)-E_\text{TDC}^\text{cons}(\boldsymbol{\lambda}_\text{TDC})$.

\subsection{Outage Probability of Offloading}

In this section, we investigate the outage probability of offloading workload to the data center-enabled high-altitude platform (HAP). Specifically, we analyze the scenario where the established transmission link between the terrestrial data center and the HAP cannot support the offloaded workload. As we assume that this link uses MIMO technology, we refer to the outage probability analysis of MIMO Rician fading channels presented in \cite{zhu2009mutual}. In \cite{zhu2009mutual}, the authors derive a lower bound and an upper bound of the outage probability's distribution by considering the trace of a non-central Wishart matrix derived from the channel matrix. Therefore, the upper bound $\text{Pr}_\text{tr\_U}$ of the complementary cumulative distribution function (CCDF) of the data rate for the transmission link to the HAP is given by \cite{zhu2009mutual}:
\begin{equation}
\label{eq:UB}
\text{Pr}\left(R_{\text{HAP}}>\lambda\right) \leq \text{Pr}_\text{tr\_U}(\lambda)\!= \!Q_{MN}\!\left(\!\sqrt{2\delta'},\sqrt{2aN\!\left(\!2^{\frac{\lambda}{N}}\!-\!1\!\right)}\!\right),
\end{equation}
where $\lambda =\sum \boldsymbol{\lambda}_\text{HAP}$ is the total offloaded workload,  $\delta'=\zeta \norm{\overline{\boldsymbol{H}}_{\text{HAP}}}_F^2$ is the non-centrality parameter, $a=\frac{1+\zeta}{\eta'}$ with $\eta'$ defined as the average signal to noise ration (SNR) at each receive antenna and $Q_{MN}(a,y)$ is the generalized Marcum Q-function of order $M\times N$ given by:
\begin{equation}
    Q_{MN}(a,y)=\int_{y}^{\infty}x\left({\frac{x}{a}}\right)^{MN-1}\:e^{-\frac{(x^2+a^2)}{2}}\:I_{MN-1}(ax)dx,
\end{equation}
with $I_{MN-1}$ is denoting the modified Bessel function of the first kind of order $MN-1$. 

Moreover, the lower bound $\text{Pr}_\text{tr\_L}$ of the CCDF of the HAP data rate is given by \cite{zhu2009mutual}:
\begin{equation}
\label{eq:LB}
\text{Pr}(R_{\text{HAP}}>\lambda) \geq  \text{Pr}_\text{tr\_L}(\lambda)=  Q_{MN}\left(\sqrt{2\delta'},\sqrt{2a\left(2^{\lambda}-1\right)}\right).
\end{equation}

Given the upper bound and the lower bound of the CCDF of the HAP data rate, we investigate the dropping rate of the data center-enabled HAP; which is given by:
\begin{equation}
   \text{Pr}_\text{drop}(\lambda)=1-\text{Pr}(R_{\text{HAP}}>\lambda). 
\end{equation}
% Therefore, the dropped workload $\lambda^\text{drop}=\lambda\: \text{Pr}_\text{drop}(\lambda)$ can be bounded as follows:
 %\begin{align}
% &\lambda\:\left(1-Q_{MN}\left(\sqrt{2\delta'},\sqrt{2a'N\left(2^{\frac{\lambda}{N}}-1\right)}\right)\right)   \leq \lambda^\text{drop}\\ 
% &\leq \lambda\:\left(1-Q_{MN}\left(\sqrt{2\delta'},\sqrt{2a'\left(2^{\lambda}-1\right)}\right)\right).\nonumber
% \end{align}

 To mitigate this dropping rate,  the dropped workload can be re-transmitted thanks to the energy saved by the data center-enabled HAP (as studied in the previous sub-section). Accordingly, the number of re-transmissions $N_\mathrm{r}$ can be expressed as a function of the dropped workload $\lambda^\text{drop}$ as follows: 
  \begin{align}
& \quad N_\mathrm{r}  =
    \begin{cases} 
            \lceil \frac{E^\text{sav}(\lambda)}{E_\text{TDC-HAP}^\text{trans}(\lambda^\text{drop})}\rceil, &         \text{if } \lambda \geq \lambda^\text{drop*},\\
            0, &         \text{otherwise}.
    \end{cases}
    \label{eq:Nr}
 \end{align}
 with $\lambda^\text{drop*}$ denotes the maximum workload arrival with no dropped workload such that $\text{Pr}(R_{\text{HAP}}>\lambda)=1,\: \forall \lambda \leq \lambda^\text{drop*}$. % By using the bounds of the dropping rate, we can determine the maximum number of possible re-transmissions $Nr^\text{max}$ as follows:
%\begin{align}
%& \quad Nr^\text{max}= \begin{cases} 
 %\lceil \frac{E^\text{sav}(\lambda)}{E_\text{TDC-HAP}^\text{trans}\left(\lambda\:\left(1-Q_{MN}\left(\sqrt{2\delta'},\sqrt{2a'N\left(2^{\frac{\lambda}{N}}-1\right)}\right)\right)\right)}\rceil,          \text{if } \lambda \geq \lambda^\text{drop*},\\
 %           0, \quad         \text{otherwise}.
 %   \end{cases}
%\end{align}

%$\lambda_\text{min}^\text{drop}$ denote the minimum workload arrival rate in the upper bound  where all the packets are dropped such that $\text{Pr}(R_{\text{HAP}}>\lambda)=0,\: \forall \lambda \geq \lambda_\text{min}^\text{drop}$. 

\subsection{Delay in Data Center-enabled HAP}
In this section, we analyze the delay $\mathcal{D}$ that occurs in a data center-enabled high-altitude platform (HAP) system. This delay is primarily composed of two components: the waiting time, also known as the queuing time, denoted as ${\mathcal{W}_i}$, for each task to be executed on server $s_i$, and the round trip time (RTT) required to send the task to the HAP and receive the execution result, such that $\mathcal{D}=\overline{\mathcal{W}_i}+{\mathcal{RTT}}$. The RTT is determined by the transmission delay and can be expressed as follows:
 \begin{equation}
    \mathcal{RTT}(l,\boldsymbol{\lambda}_\text{HAP})=2\: t_\text{trans}(l,\boldsymbol{\lambda}_\text{HAP})= \frac{2\:b(\boldsymbol{\lambda}_\text{HAP})}{R_{\text{HAP}}(l)}. 
 \end{equation}

In order to calculate the mean waiting time, we model the server as an M/G/1 queue with vacations, since the server enters an idle mode when there are no tasks to be executed. Our approach closely follows the standard derivations for M/G/1 queues, as described in \cite{adan2015department, xia2012max, elwhishi2012self}. Specifically, we employ the following:
\begin{itemize}
    \item the arriving workload follows a Poisson process with rate $\lambda_{i}$.
    \item the service time distribution is general because the control commands are assumed to be random. %with service time $\mu=\frac{{C}_{Ter}}{\text{packet size}}$
\end{itemize}

 We are interested in obtaining the mean waiting time $\overline{\mathcal{W}_i}$ at server $s_i$ and its second moment $\overline{\mathcal{W}_i^2}$. The waiting time  of task $k$ scheduled after $K'$ tasks  is  given by ${\mathcal{W}_i}=\sum_{k'=1}^{K'}{\mathcal{X}_i^{k'}}+\mathcal{R}_i$; 
%\begin{equation}
 %   \mathcal{W}_i=\sum_{k'=1}^{K'}{\mathcal{X}_i^{k'}}+\mathcal{R}_i\; ,
 %   \label{eq:wait}
%\end{equation}
where   $\mathcal{X}_i^{k'}$ is the service time of task $k'$  that arrived before task $k$ at server $s_i$ and $\mathcal{R}_i$ is the residual service time of server $s_i$. $\mathcal{R}_i$ can be either residual service time $\mathcal{R}_i^\mathrm{s}$ or residual
vacation time $\mathcal{R}_i^\mathrm{v}$ depending on utilization. The first moment
of this residual time can then be written as:
\begin{align}
  \label{eq:resid_time}
   \overline{\mathcal{R}_i}&= {\overline{\mathcal{R}_i^\mathrm{s}}+\overline{\mathcal{R}_i^\mathrm{v}}}=\frac{1}{2}\left(u_i\:\frac{\overline{X_i^2}}{\overline{X_i}}+(1-u_i)\:\frac{\overline{V_i^2}}{\overline{V_i}}\right)\\
   &=\frac{1}{2}\left(\lambda_i\:\overline{X_i^2}+(1-u_i)\:\frac{\overline{V_i^2}}{\overline{V_i}}\right)\nonumber;
\end{align}
where $u_i$ is the utilization ratio of server $s_i$, $\overline{V_i}$ is the mean vacation time duration at server $i$ and $\overline{V^2_i}$ is the second moment of vacation time duration at server $i$. 
By applying Little's formula, we obtain the average waiting time as follows: %$ \overline{\mathcal{W}_i}=\frac{\overline{\mathcal{R}_i}}{1-u_i}$.
\begin{equation}
\overline{\mathcal{W}_i}=\frac{\overline{\mathcal{R}_i}}{1-u_i}=\frac{\lambda_i\:\overline{X_i^2}}{2\:(1-u_i)}+\frac{\overline{V_i^2}}{2\:\overline{V_i}}.
\label{eq:mean_wait}
\end{equation}
%Therefore, the second moment of the waiting time as follows:
%\begin{align}
%\overline{\mathcal{W}_i^2}&=\lambda_i\:\overline{\mathcal{W}_i}\:\text{Var}(\mathcal{X})+\left(1+\frac{u_i}{1-u_i}\right)^2\overline{\mathcal{R}_i}^2+\text{Var}(\mathcal{R}_i)\\
%&=\lambda_i\:\overline{\mathcal{W}_i}\:\text{Var}(\mathcal{X})+\frac{u_i^2}{(1-u_i)^2}\:\overline{\mathcal{R}_i}^2+\overline{\mathcal{R}_i^2}\:;
%\end{align}
%where $\overline{\mathcal{R}_i^2}$ is obtained by applying  {the law of total
%expectation}; which yields:
%\begin{align}
 %  \overline{\mathcal{R}_i^2}&=  {\overline{(\mathcal{R}_i^\mathrm{s})^2}+\overline{(\mathcal{R}_i^\mathrm{v})^2}}
   % =\frac{1}{3}\left(u_i\:\frac{\overline{X_i^3}}{\overline{X_i}}+(1-u_i)\:\frac{\overline{V_i^3}}{\overline{V_i}}\right)\\
    %&=\frac{1}{3}\left(\lambda_i\:{\overline{X_i^3}}+(1-u_i)\:\frac{\overline{V_i^3}}{\overline{V_i}}\right)\nonumber.
%\end{align}

 {
\subsection{Complexity Analysis}
In this subsection, we analyze the complexity of the proposed data center enabled HAP system and we discuss its feasibility for implementation in a realistic setup. The complexity analysis considers various factors that include the computational complexity, the communication complexity and the deployment complexity. 
\subsubsection{Computational Complexity}
To assess the computational complexity of the proposed framework, we consider the time and resources required to process the offloaded workload form the terrestrial data center.  The  energy saving complexity of our proposed framework can be evaluated as $\mathcal{O}(\dfrac{\lambda \times N}{\mu})$ where $\lambda$ is the workload arrival rate $N$ is the total number of servers in the system and $\mu$ is the server capability. The time complexity of our proposed framework can be evaluated as $\mathcal{O}(\dfrac{\lambda \times N}{\mu^2})$; where we consider the waiting time and the offloading time of the tasks to the HAP.
We notice the trade-off between the energy saving offered by the HAP-enabled system and the increased computational requirements due to the limited resources available on the HAP.
Specifically, the more servers are deployed in the HAP with increasing workload offloading, the more energy is saved in the terrestrial data center. However, this strategy impacts resource utilization since energy consumption is strongly coupled to servers' capabilities.
Accordingly, over-utilization and under-utilization of the flying servers must be avoided during resource' provisioning and allocation \cite{abderrahim2023how}. 
\subsubsection{Communication Complexity}
The communication complexity of the proposed framework involves the analysis of the communication links between the terrestrial data centers and the HAP. This includes the evaluation of the bandwidth requirements, signal propagation delays and the reliability of these links through the outage probability as studied in section 3.3. We can model the communication complexity by using a combination of metrics such as the bit error rate (BER), signal-to-noise ratio (SNR) and throughput.
\subsubsection{Deployment Complexity}
The deployment complexity of the proposed framework involves the practical challenges associated with deploying and maintaining HAPs in the stratosphere. This complexity encompasses the  HAP positioning and buoyancy control; which are tightly related to weather conditions in the stratosphere determined mainly by the wind speed. Moreover, this complexity includes the use of sophisticated and resilient electronic devices in the flying servers to guarantee their operation in the low temperature of the stratosphere and afford a reliable computing service.  The deployment complexity involves also the regulatory rules; which impose strict conformity with the standards of design, transport and operation to guarantee airships' safety in the stratosphere.\newline\\
In summary, the complexity analysis of the proposed framework demonstrates its feasibility for implementation in a realistic setup while considering the important trade-offs. The energy savings and environmental benefits offered by the data center enabled HAP outweigh the computational and communication complexity. Moreover, the deployment challenges can be addressed through technological advancements and regulatory efforts. 
}
%%%%%%%%%%%%%%%%%%%%%%%%%%%%%%%%%%%%%%%%%%%%%%%%%%%%%%%%%%%%%%%%%%
\section{Results and Discussion}

In this section, we consider a stratospheric airship-based high-altitude platform (HAP) that has a significant surface area and can harvest notable amounts of solar energy. We assume that the HAP can support a maximum payload of $450 \text{kg}$, similar to the Stratobus airship HAP \cite{strat,kurt2021vision}. Additionally, we assume that the hosted data center has a rack weight of $363 \text{kg}$ \cite{micoDC}, with each server in the rack characterized by a service rate of $\mu=580\text{MIPS}$ and a weight of approximately $9 \text{kg}$ \cite{micoDC}. The remaining parameters used in our simulations are detailed in Table \ref{table:2}.

Throughout our numerical results, we employ the derived expressions in section 3 to quantify the performance gain of the data center-enabled HAP compared to a terrestrial data center under various operational conditions.  {To analyze the scalability of our solution, we consider various workload scenarios including the homogeneous workload and heterogeneous workload. The considered workload model in each scenario has a wide range of arrival rates throughout all the conducted simulations. The homogeneous workload scenario involves workloads with the same characteristics in terms of task's size. The heterogeneous workloads scenario involves workloads characterized with small task length and workloads characterized with large task length. For each workload scenario, we investigate the data center enabled HAP's ability to handle the increased demand by augmenting the workload arrival rate without significantly impacting the performance metric.}

 {We consider four simulation scenarios to investigate the impact of HAP location over the course of a year, workload arrival rate, and the number of servers present in the HAP on the relative performance gain. Firstly, we verify the HAP's flying condition in the homogeneous workload scenario by monitoring the energy balance based on the number of servers in the HAP and the maximum workload arrival rate. Secondly, we evaluate the energy efficiency of the data center-enabled HAP by assessing the saved energy for different latitudes throughout the year. To further validate the scalability of our proposed framework, we consider the deployment of a constellation of HAPs along with the terrestrial data center. Thirdly, we explore the impact of re-transmitting dropped tasks in case of link outage on system performance for heterogeneous workload scenario. Finally, we analyze the delay experienced in the data center-enabled HAP by comparing the queuing delay, i.e., the waiting time, to the transmission delay in the heterogeneous workload scenario.}

 \begin{table}[H]
\caption{ Simulation Settings}
\centering
\begin{tabular}{|c|c |c|} 
 \hline
Type &Parameter & Numerical Value  \\ 
 \hline\hline
%Computing & Server's Idle Power \cite{zhabelova2018comprehensive} & $5$ kW\\
%Inputs & Server Peak Power \cite{zhabelova2018comprehensive} & $10$ kW\\
%&MIPS \cite{li2017holistic} &  500 \\
% \hline
& Supply Temperature  & $299.15$ K\\
Cooling/& Server Initial Temperature  & $310$ K\\
Thermal &CPU Initial Temperature  & $318$ K\\
Inputs & Thermal Resistance & $0.34$ K/W\\
\cite{li2017holistic} & Server Heat Capacity  & $340$  J/K\\
 \hline
  & Area of the PV   & $8000\:\text{m}^2$ \\
 HAP &Efficiency of the PV   & $0.4$ \\
Inputs &Propeller efficiency  & $0.8$ \\
\cite{sun2020simulation,arum2020energy} &Air Density  & $0.08891\: kg/m^3$ \\
 &Dynamic Air Viscosity & $1.422.10^{-5}\: N.s/m^2$ \\
 \hline
 Trans-& Antennas in TDC  & 2  \\
mission &Antennas  in HAP  & 16  \\
Inputs &Carrier Frequency  & $31$ GHz \\
\cite{ding2021joint}&Channel Bandwidth   & $100$ MHz \\
 \hline 
\end{tabular}
\label{table:2}
\end{table}

 \subsection{HAP Flying Condition}
 \begin{figure}
     \centering
     \includegraphics[width=3.5in]{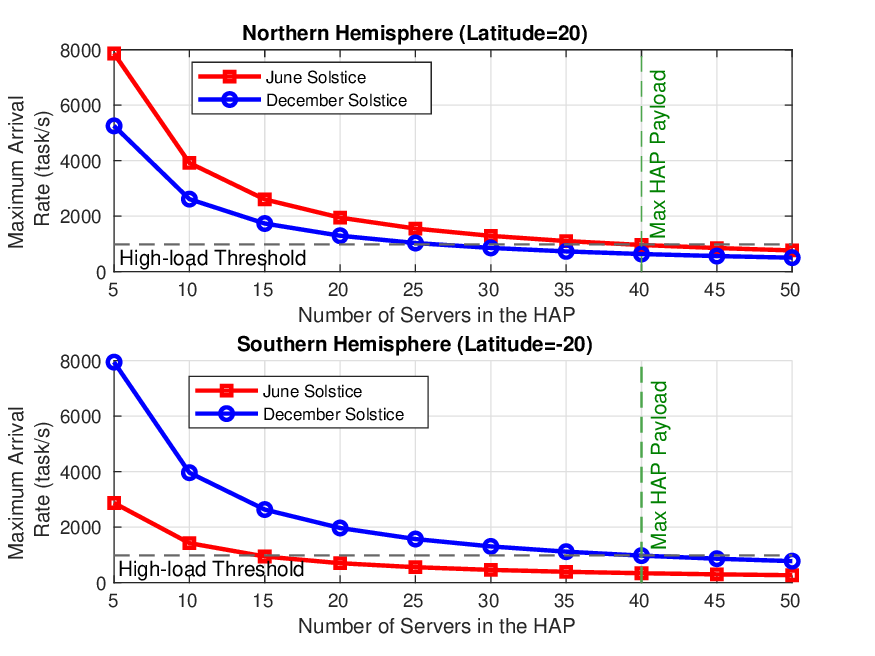}
     \caption{Maximum Workload Arrival Rate under the HAP Flying Condition}
     \label{fig:nb_Serv1}
 \end{figure}
 
 The first simulation example investigates the workload arrival rate, i.e., $\lambda^\text{max}$, that maintains the HAP flying condition versus the number of HAP servers, i.e., $I'$, for different locations and time along the year, as depicted in Fig.~\ref{fig:nb_Serv1}. It is necessary to consider the physical capacity of our system beforehand. Specifically, the maximum payload supported by the HAP  imposes a limit on the maximum number of servers hosted in the HAP, as shown in Fig.~\ref{fig:nb_Serv1}. Moreover, the server's service rate characteristic $\mu$ combined with the desired utilization ratio $\overline{u}$ dictate a limit on the supported workload. Since we want to leverage the full performance of the HAP, we assume that the HAP is fully occupied and consider the Maximum HAP Payload limit in our simulations. Moreover, we suppose that the servers hosted in the HAP are fully-utilized by imposing that $\overline{u}\approx 100\%$. This maximum utilization ratio yields a High-load Threshold for the workload arrival rate, as shown in Fig.~\ref{fig:nb_Serv1}; beyond which the servers are dysfunctional.
 
We observe first from Fig.~\ref{fig:nb_Serv1} that the harvested energy is under-utilized when the number of airborne servers is low because the desirable utilization ratio $\overline{u}$ restricts the accepted workload. For instance, the supported maximum arrival rate in the Northern Hemisphere is around $2600$ task/s for ten servers on December Solstice. However, the actual server capacity of the server is below $1000$ task/s even when it is highly loaded. Therefore, the number of servers in the HAP should be increased to utilize the harvested energy fully. However, there is a limitation on the maximum number of airborne servers according to the supported HAP payload. Indeed, we notice that the harvested energy and the servers' capacity are ideally utilized when the number of servers reaches the maximum HAP payload. Moreover, we notice a lower maximum arrival rate allowed per server when more servers are present in the HAP. This observation is because the same number of tasks can be distributed and processed by more servers to maintain the flying condition. Otherwise, the harvested energy would not cover the payload energy, which is equivalent to the computational energy of the servers. We also notice an opposite behavior for the maximum arrival rate when comparing the Northern hemisphere to the Southern hemisphere. For instance, in the Northern hemisphere, more tasks are accepted during June than December solstice. However, more tasks are carried out in the Southern hemisphere during December than June solstice. This observation is because the solar radiation and the daylight duration are more critical during June in the Northern hemisphere; because the Northern hemisphere is closer to the Sun during June. However, the solar radiation and the daylight duration are more critical during December in the Southern hemisphere because the Southern hemisphere is closer to the Sun during December. Therefore, more solar energy can be harvested during June in the Northern hemisphere and December in the Southern hemisphere. Hence, more computational energy can be covered, and accordingly, more tasks can be processed in the HAP during June in the Northern hemisphere and December in the Southern hemisphere. 
 
\begin{figure}
     \centering
     \includegraphics[width=3.5in]{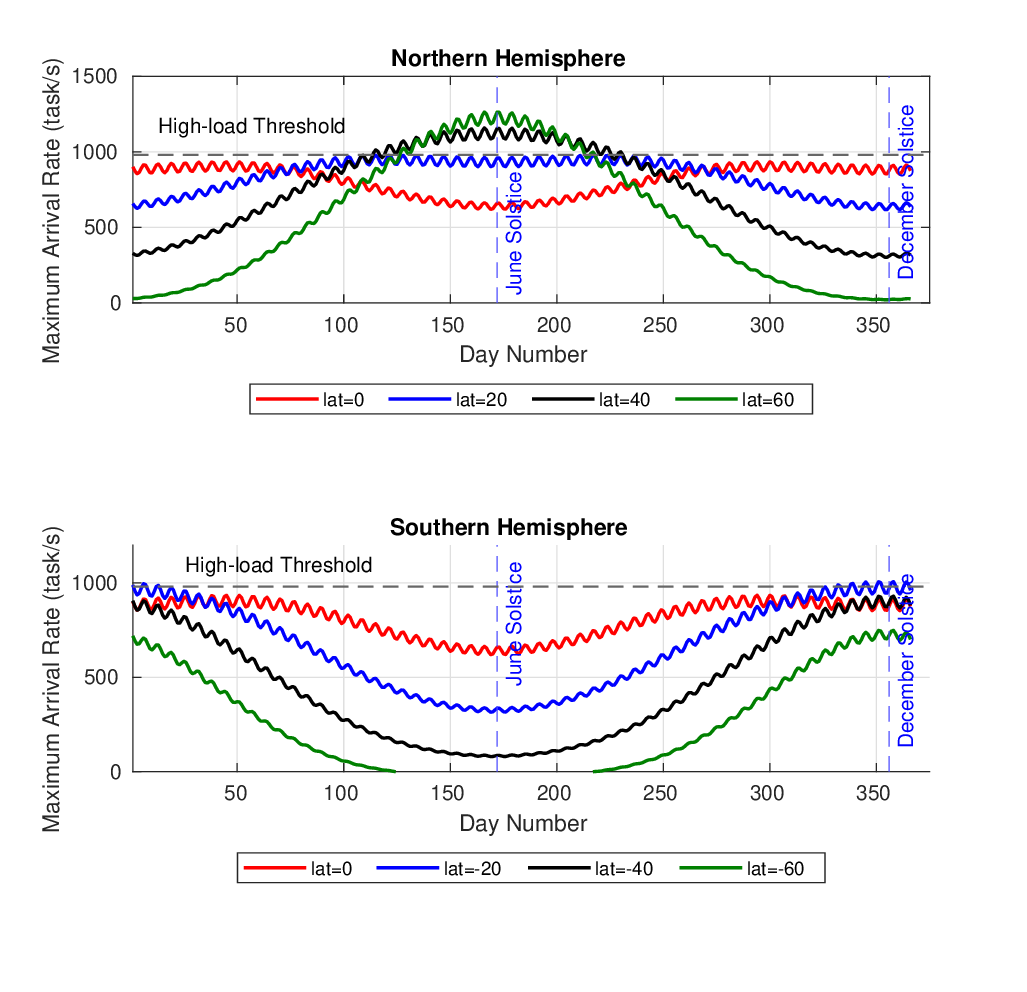}
     \caption{Maximum Workload Arrival Rate under the HAP Flying Condition. ($I'=40$)}
     \label{fig:day1}
 \end{figure}

Since HAPs are designed to operate for long-duration missions, we assume that the full payload is used when the airship is launched, and we study the maximum workload arrival rate variation according to days and latitudes. We note that it is crucial to consider the High-load Threshold imposed by the desirable utilization ratio of the servers. Indeed, as depicted in Fig.~\ref{fig:day1}, the servers tend to be over-utilized around June solstice in the Northern hemisphere and under-utilized in the Southern hemisphere if the utilization ratio is overlooked. More generally, the servers' computational capacity is better utilized with higher latitudes in both hemispheres. 
Moreover, we notice that the maximum arrival rate reaches on June solstice its maximum point in the Northern hemisphere and its minimum point in the Southern hemisphere. The explanation of this observation was deeply detailed in the previous paragraph. We notice also that the maximum arrival rate increases with latitudes in the Southern hemisphere. In the Northern hemisphere, the maximum arrival rate decreases with latitudes when $d\in [1,100]\cup[250,366]$ and increases with latitudes when $d\in[100,250]$.  It is worthy to mention that the zigzag course of the curves is due to the expression of the solar radiation intensity given in (\ref{eq:rad_int}).   We notice also that the maximum arrival rate has less variations around the equator compared to the other latitudes along the year.  Indeed, for high latitudes, the maximum arrival rate is less dynamic and is around the full capacity of the servers during different periods of the year. This observation is important because it shed light into  the workload management along the year given the long-duration of HAP's missions. 

%In conclusion, the workload arrival rate scheduled to the servers in the HAP should be studied based on the considered day and latitude. To maintain the flying condition of the HAP, the harvested solar energy should cover the energy requirements of the data center-enabled HAP. Therefore, the workload arrival rate scheduled to the HAP $\boldsymbol{\lambda}_\text{HAP}$ must satisfy the condition $\boldsymbol{\lambda}_\text{HAP} < \boldsymbol{\lambda}_\text{HAP}^\text{max}$. %We assume that the HAP can drop the workload that exceeds $\boldsymbol{\lambda}_\text{HAP}^\text{max}$ to maintain the flying condition. 

  \subsection{Energy Saving}
\begin{figure}
    \centering
    \includegraphics[width=3.5in]{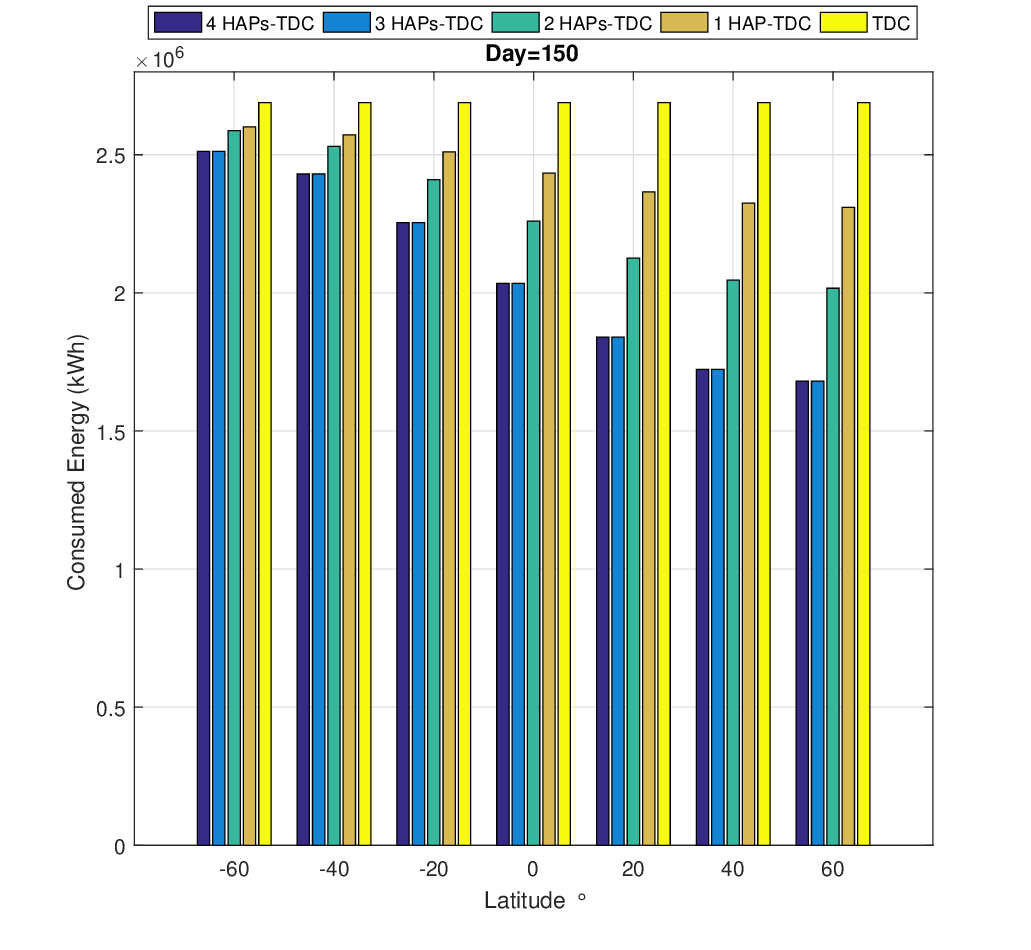}
    \caption{Energy consumption according to Latitude's variation }
    \label{fig:energy1}
\end{figure}
After checking the HAP flying condition and determining the maximum arrival rate for different days and latitudes, we explore the energy-saving capabilities of the data center-enabled HAP. This simulation evaluates the consumed energy for the maximum workload arrival rate relative to different latitudes and days. First, we study the consumed energy variation for both data center-enabled HAP and terrestrial data centers versus the latitude for day number $150$ in the year, as depicted in Fig.~\ref{fig:energy1}. We observe first that the consumed energy decreases with latitude because more workload can be accepted and hence covered by the HAP's harvested energy. We also notice that the data center-enabled HAP helps to reduce the consumed energy for a large range of latitude $[-60^\circ,60^\circ]$. We assess the saved energy rate  $E^\text{sav}(\%)=14.61\%$ achieved for the maximum workload arrival rate $\lambda^\text{max}$ when the latitude is around $60^\circ$ for one HAP. Higher energy saving rates can be reached with a lower arrival rate $\lambda<\lambda^\text{max}$. We also note that if a larger payload can be supported by the HAP (increased by ten servers), higher energy saving can be recorded (increased by $3\%$). Indeed, a more important workload amount can be processed in the HAP in this case, which reduces the cooling energy consumption and yields higher energy savings.
\begin{figure}
    \centering
    \includegraphics[width=3.5in]{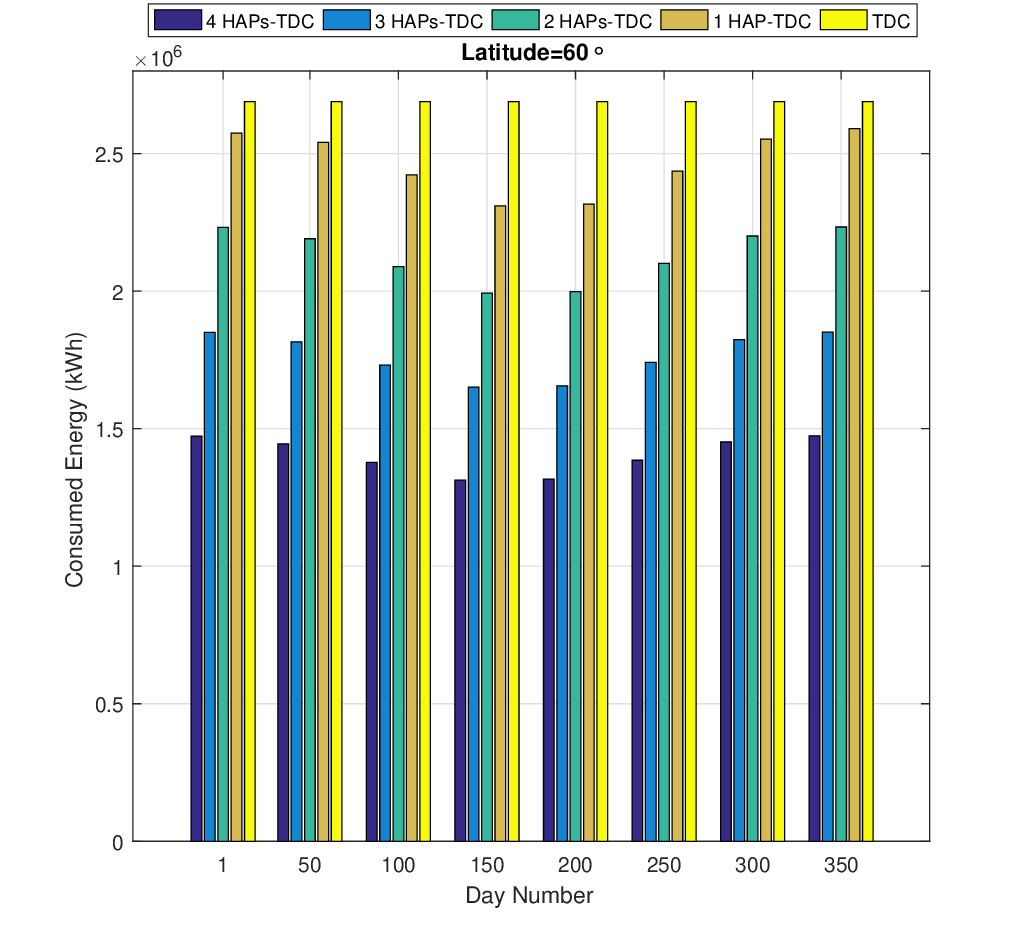}
    \caption{Energy consumption according to Day's variation}
    \label{fig:energy2}
\end{figure}

 Then, we study the consumed energy variation according to different days in the year, as shown in Fig.~\ref{fig:energy2}. The least consumed energy record is at the middle of the year for the data center-enabled HAP. This observation is because the beginning and the end of the year are characterized by cold weather (the Winter season), where the ambient temperature reaches its lowest levels. Hence, less solar energy can be harvested by the HAP. However, the days around $200^\text{th}$ day belong to Summer; where more solar energy can be collected within the data center-enabled HAP. We also notice that the data center-enabled HAP helps to reduce the consumed energy with an energy-saving rate $E^\text{sav}(\%)=14.38\%$ achieved for the maximum workload arrival rate $\lambda^\text{max}$ around the $200^\text{th}$ of the year. These saving rates can be substantially enhanced with the deployment of more HAPs. For instance,  the saving rate exceeds $17\%$ starting from 2 HAPs deployment along with the terrestrial data center.

\subsection{Outage Probability of Offloading}
\begin{figure}
    \centering
    \includegraphics[width=3.5in]{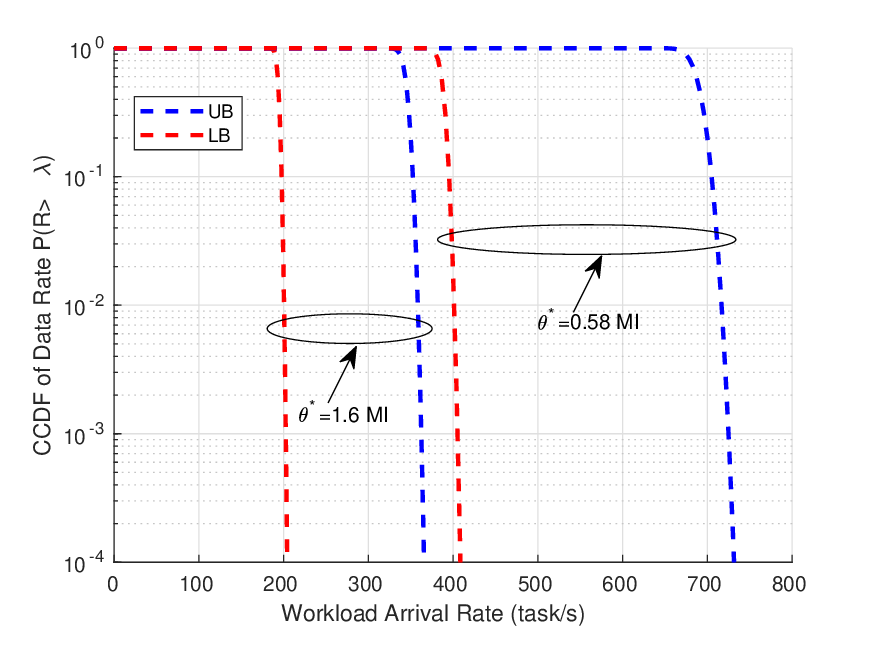}
    \caption{Outage Probability of the Arriving Workload}
     \label{fig:out1}
\end{figure}

This simulation studies the outage probability of the communication links between terrestrial and HAP-based data centers. To this end, we plot the upper bound (\ref{eq:UB}) and lower bound (\ref{eq:LB}) outage probabilities  respectively, versus the workload arrival rates, as illustrated in Fig.~\ref{fig:out1}. We note that the partial dropping of some tasks might start at a workload arrival rate around $\lambda_\text{partial}^\text{LB}=360\: \text{task/s}$ by considering the lower bound and at a workload arrival rate around $\lambda_\text{partial}^\text{UB}=650\: \text{task/s}$ by considering the upper bound. Hence,  $\lambda^\text{drop*}=360\: \text{task/s}$ (c.f. (\ref{eq:Nr})).  The total dropping of the arriving tasks starts at  a workload arrival rate around $\lambda_\text{total}^\text{LB}=400\: \text{task/s}$ by considering the lower bound and at a workload arrival rate around $\lambda_\text{total}^\text{UB}=730\: \text{task/s}$ by considering the upper bound. We also study the impact of the mean task length on the outage probability. As depicted in Fig.~\ref{fig:out1}, we notice that the outage probability is more important for a higher task length because of more data ($b$ in bits (\ref{eq:E_trans})) is carried on the transmission link for the same workload arrival rate.  

\begin{figure}
    \centering
    \includegraphics[width=3.5in]{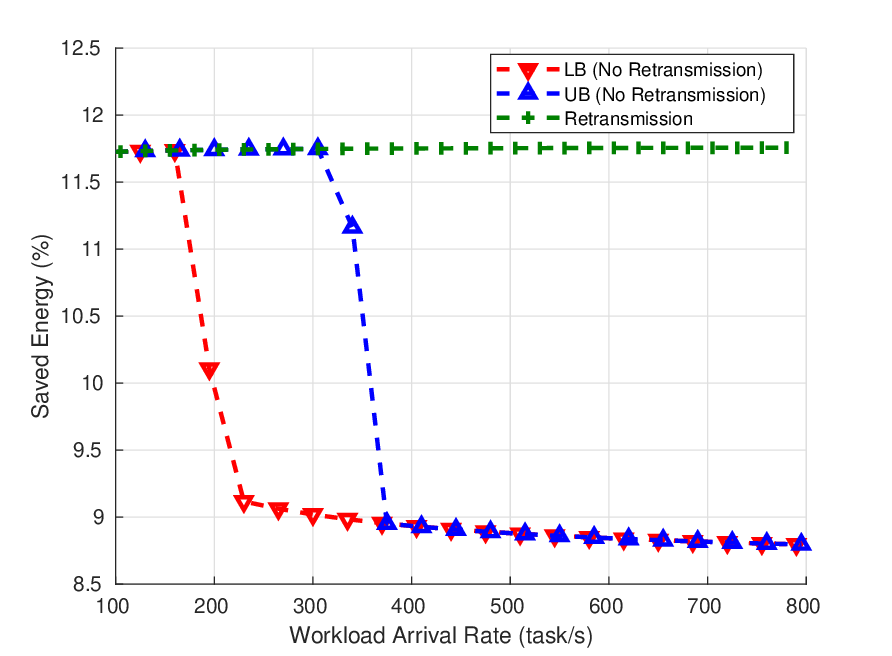}
    \caption{Saved Energy  in the case of Re-transmission to HAP}
    \label{fig:out2}
\end{figure}

Given the outage probability and the saved energy findings, we can investigate the tasks' retransmission impacts on the saved energy to the HAP. Therefore, we study the saved energy variation according to the workload arrival rate in two cases. In the first case, the dropped workload is processed in the terrestrial data center, while in the second case, we consider that the dropped workload is re-transmitted to the HAP. As shown in Fig.~\ref{fig:out2}, if the dropped workload is not re-transmitted to the HAP, the saved energy percentage decreases notably with high arrival rates. Interestingly, the saved energy through a data center-enabled HAP is around $9\%$ even when the workload is partially processed in the flying data center. However, under the re-transmission assumption, the saved energy becomes around $11.76\%$, which achieves almost the same performance in the case of zero outage. This observation highlights that the saved energy is not impacted even for high data rates because the transmission energy is significantly lower than the cooling energy. 

\subsection{Delay in Data Center-enabled HAP}
\begin{figure}
      \centering
      \includegraphics[width=3.5in]{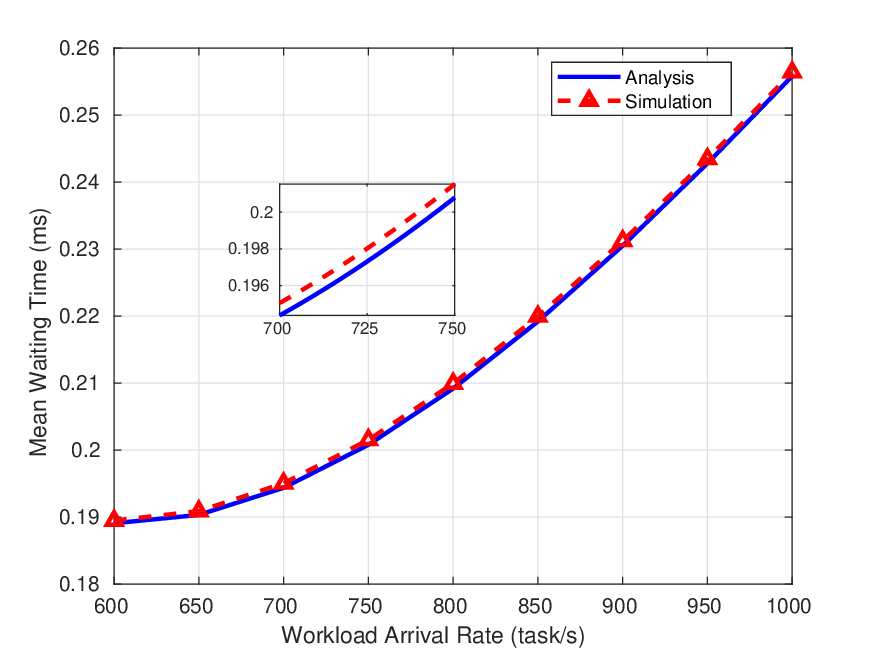}
      \caption{Mean Waiting Time Variation according to the Arriving Workload}
      \label{fig:wait}
  \end{figure}

In this simulation, we study the mean waiting time (\ref{eq:mean_wait}) in our system.  Therefore, we approximate the general distribution of the considered queuing model to an exponential service time distribution with a mean service rate ${\mu_\mathrm{s}}$ and an exponential vacation time distribution with a mean vacation rate ${\mu_\mathrm{v}}$. Accordingly, the first moment of the residual service time can be simplified, based on (\ref{eq:resid_time}), as   follows:
\begin{equation}
\overline{\mathcal{R}_i}=\frac{\lambda_i}{\mu_\mathrm{s}^2}+\frac{(1-u_i)}{\mu_\mathrm{v}}\;.
\end{equation}
By using this approximation, we compare the mean waiting time obtained through the analysis in section 3.4 to the simulation results of an M/M/1 queue model with vacations where the service rate is ${\mu_\mathrm{s}}$ and the vacation rate is ${\mu_\mathrm{v}}$. Therefore, we study the mean waiting time according to the variation of the workload arrival rate in both cases.  As depicted in Fig.~\ref{fig:wait}, the analytical results coincide with the simulation results for different workload arrival rates. Indeed, the difference between both curves $|\overline{W_i}^\text{analysis}-\overline{W_i}^\text{simulation}|\leq 10^{-6}$ as zoomed out for the values of workload arrival rates around 700 task/s. We notice also that higher delays are attained when the workload arrival rate increases because more tasks should be waiting before their processing.   

\begin{figure}
    \centering
    \includegraphics[width=3.5in]{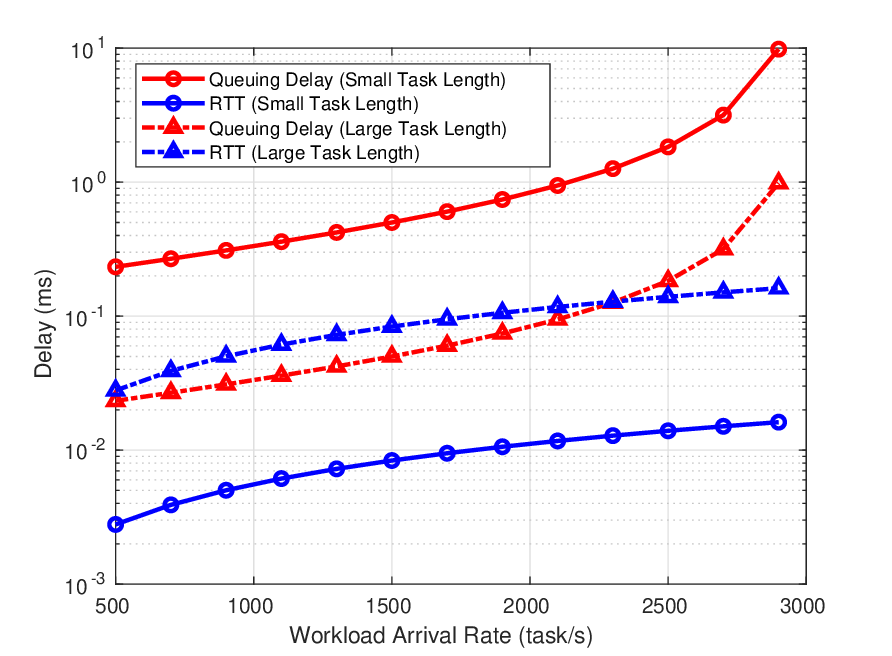}
    \caption{Comparison between the Queuing Delay and the Transmission Delay }
    \label{fig:compar_delay}
\end{figure}

Therefore, we study the experienced delay in a data center-enabled HAP according to the variation of the workload arrival rate. As depicted in Fig.~\ref{fig:compar_delay}, we notice that this comparison depends on the mean task length of the arriving workload. For instance, the queuing delay is notably more significant than RTT  for different arrival rates in the case of small task length. These results indicate that the tasks' transmission and the re-transmission to the HAP can occur without affecting the experienced delay for a workload characterized by a small task length. However, the queuing delay is sometimes lower than RTT for a given range of arrival rates in the case of large task length.   These results indicate that the tasks' transmission and re-transmission to the HAP  should be carefully studied for a workload characterized with a large task length because it depends on the arrival rate. For example, the queuing delay is close to RTT for an arrival rate $500\:\text{task/s}\leq\lambda\leq2250\:\text{task/s}$  as depicted in Fig.~\ref{fig:compar_delay}. Therefore, the tasks' transmission to the HAP will add an additional and substantial delay to the queuing delay in this case. But, the queuing delay becomes higher than RTT for an arrival rate $2250\:\text{task/s}\leq\lambda\leq3000\:\text{task/s}$ as depicted in Fig.~\ref{fig:compar_delay}. We note that the difference is not important, though, and would limit the number of re-transmitted tasks in the case of an outage.

\vspace{-0.3cm}
\section{Conclusion}
Throughout this study, we have explored the potential of using a data center-enabled HAP system as a green alternative to traditional terrestrial data centers. Our analysis shows that the naturally low temperature and solar power available in the stratosphere make the HAP an ideal environment for flying data centers. We have established the limits of our system in terms of payload capacity and server utilization ratio to determine the optimal workload arrival rate to the HAP.
Our study also demonstrates the energy-saving benefits of using a data center-enabled HAP compared to traditional terrestrial data centers under various workload settings, locations, and periods of the year. We have found that energy savings can be further improved by reducing the workload arrival rate to the HAP or increasing the number of servers hosted in the HAP. To increase system reliability and save more energy, reliable communication links are necessary to re-transmit dropped workloads to the HAP.
Additionally, we have shown that the distribution of servers between terrestrial data centers and the data center-enabled HAP has no significant impact on the transmission and re-transmission delay for workloads with low task length. However, careful design of workload scheduling is essential to take full advantage of the benefits offered by the HAP system.
Overall, our findings suggest that a data center-enabled HAP system presents a promising solution for reducing energy consumption and mitigating the environmental impact of traditional terrestrial data centers. Our study provides valuable insights into the potential benefits of this innovative computing paradigm and highlights the need for continued research in this area.
 {
\subsection{Research Challenges}However, it is important to explore  the potential research challenges towards a comprehensive study. Therefore, the technical challenges of the data center enabled-HAP should be  investigated. For instance, the unfriendly weather conditions in the stratosphere impose the deployment of adapted and more resilient electronic devices in the flying data center. Also, the frequency's maintenance of the airship and the flying servers is crucial to balance the trade-off between the quality of the offered computing service and the HAP's mission duration. Moreover, the economic viability of the data center enabled HAP should be considered. Capital expenditures such as the costs of the HAP platform and aerial servers, as well as operational expenditures like energy costs, must be assessed to evaluate profitability. Accordingly, a dynamic computing pricing model for the HAP is needed. Prices should adjust based on demand and network conditions to provide a satisfactory quality of experience for users while maximizing utilization of HAP resources and ensuring profitability \cite{mitsis2022price}. Specifically, lower prices can be initially offered to encourage offloading to the HAP during off-peak periods. However, as the arrival rate approaches the HAP's capacity, prices should increase to throttle demand or incentivize offloading some traffic to terrestrial data centers, which can offer lower prices during peak periods.  }
 {\subsection{Future Directions}To tackle these challenges and improve the performance of the data center enabled-HAP, our future work will be based on different approaches of machine learning.  For instance,  meta learning can be useful to optimize the workload/network management policies in  the flying data center hosted in the HAP. One way to boost the self-organization of the data center-enabled HAP network is to adopt meta-learning. Meta learning can be used in the data center-enabled HAP to learn from the output of the machine learning algorithms commonly used to optimize the network management policies in the terrestrial data center. Then, the generated models can be applied in the flying data center hosted in the HAP. These advantages are particularly valuable in the HAP because the learning process is accelerated and hence more computational energy is saved.
Moreover, it is valuable to predict the highly-dynamic workload arriving at the data center by using federated learning. Federated learning can play a crucial role in predicting the workload incoming to the data center in the briefest delays without violating data privacy and less exchanging plain text data to ensure security. Indeed, federated learning can be applied within the different servers of one data center, geographically distributed sites of the same data center, or even within other data centers by only sending the prediction results and models.  Accordingly, the predicted workload should be effectively scheduled spatially and temporally to leverage the renewable energy. Moreover, the renewable energy usage can be alternated with the fossil energy usage in a data center enabled-HAP when the appropriate workload amount is offloaded to the HAP; while fulfilling the QoS requirements and respecting the physical capabilities of the data center enabled-HAP.
  }

\bibliographystyle{IEEEtran}

\bibliography{bib}

\begin{IEEEbiographynophoto}%[{\includegraphics[width=5in,height=1.25in,clip,keepaspectratio]{authors/wiem.png}}]
{Wiem Abderrahim} (S'14 - M'18) accomplished her undergraduate studies in electrical engineering at the Higher School of Communications of Tunis, Carthage University, Tunisia in 2013. She received her Doctoral Degree in Information and Communication Technologies from the same university in 2017. She worked as a lecturer and then as an adjunct professor at the Higher School of Communications of Tunis between 2014 and 2018. Currently, she is a postdoctoral fellow within King Abdullah University of Science and Technology (KAUST), Thuwal, Saudi Arabia. Her research interests include cloud computing, edge computing, network virtualization, software defined networking and recently satellite communications and machine learning.
\end{IEEEbiographynophoto}
%\vspace{-1cm}

\begin{IEEEbiographynophoto}%[{\includegraphics[width=1in,height=1.25in,clip,keepaspectratio]{authors/osama.png}}]
{Osama Amin} (S'07, M'11, SM'15) received his B.Sc. degree in electrical and electronic engineering from Aswan University, Egypt, in 2000, his M.Sc. degree in electrical and electronic engineering from Assiut University, Egypt, in 2004, and his Ph.D. degree in electrical and computer engineering, University of Waterloo, Canada, in 2010. In June 2012, he joined Assiut University as an assistant professor in the Electrical and Electronics Engineering Department. Currently, he is a research scientist in the CEMSE Division at KAUST, Thuwal, Makkah Province, Saudi Arabia. His general research interests lie in communication systems and signal processing for communications with special emphasis on wireless applications.
\end{IEEEbiographynophoto}
%\vspace{-1cm}

\begin{IEEEbiographynophoto}%[{\includegraphics[width=1in,height=1.25in,clip,keepaspectratio]{authors/basem.png}}]
{Basem Shihada} (SM'12) received the Ph.D. degree in computer science from the University of Waterloo, Waterloo, ON, Canada. He is an Associate and a Founding Professor with the Computer, Electrical and Mathematical Sciences Engineering Division, King Abdullah University of Science and Technology, Thuwal, Saudi Arabia. In 2009, he was appointed as a Visiting Faculty Member with the Department of Computer Science, Stanford University, Stanford, CA, USA. His current research interests include energy and resource allocation in wired and wireless networks, software-defined networking, Internet of Things, data networks, smart systems, network security, and cloud/fog computing. 
\end{IEEEbiographynophoto}

\end{document}